\RequirePackage{fix-cm}
\RequirePackage{amsmath}

\documentclass[10pt,a4paper]{article}
\usepackage{graphicx}
\usepackage{amsfonts}
\usepackage{amssymb}
\usepackage{bm}
\usepackage{hyperref}
\usepackage{multirow}
\usepackage{tabularx}
\usepackage{xcolor}
\usepackage{tikz}
\usetikzlibrary{calc,positioning}

\usepackage{amsmath, amsfonts, amssymb, amsthm, bm, bbm, mathrsfs, mathtools}
\usepackage{enumitem, booktabs}
\usepackage{caption, subcaption, multirow} 
\usepackage[title]{appendix}
\usepackage{nccmath}
\usepackage{array}

\DeclareMathOperator*{\dis}{dis}
\DeclareMathOperator*{\inn}{inn}

\newcommand{\C}{\mathbf{C}}

\newcommand{\g}{\mathbf{g}}

\usepackage[]{footmisc}
\usepackage{titlesec}

\titleformat{\section}
  {\normalfont\fontsize{12}{15}\bfseries}{\thesection}{1em}{}

\titleformat{\subsection}
  {\normalfont\fontsize{12}{15}\itshape}{\thesubsection}{1em}{}

\titleformat{\subsubsection}
  {\normalfont\fontsize{10}{15}\itshape}{\thesubsubsection}{1em}{}

\begin{document}

\begin{center}
\LARGE{Fitting the~grain orientation distribution of a~polycrystalline material conditioned on a~Laguerre tessellation}

\vspace{1.2cm}

\large{
I. Karafi\'{a}tov\'{a}\footnote[1]{email: karafiatova@karlin.mff.cuni.cz}\footnote[2]{\label{fn2}Department of Probability and Mathematical Statistics, Faculty of Mathematics and Physics, Charles University, Sokolovsk\'{a} 83, Praha 8, 186 75, Czech Republic}, J. M\o ller\footnote[3]{Department of Mathematical Sciences, Faculty of Engineering and Science, Aalborg University, Skjernvej~4A, 9220 Aalborg \O, Denmark}, Z. Pawlas\footnotemark[2],
J. Stan\v{e}k\footnote[4]{Department of Mathematics Education, Faculty of Mathematics and Physics, Charles University, Sokolovsk\'{a} 83, Praha 8, 186 75, Czech Republic},
F. Seitl\footnotemark[2],
V. Bene\v{s}\footnotemark[2]
}
\end{center}

\normalsize

\vspace{0,7cm}

\begin{abstract}
The description of distributions related to grain microstructure helps physicists to understand the~processes in materials and their properties.
This paper presents a~general statistical methodology for the~analysis of crystallographic orientations of grains in a~3D Laguerre tessellation dataset which  represents the~microstructure of a~polycrystalline material. We introduce complex stochastic models which may substitute expensive laboratory experiments:
conditional on the~Laguerre tessellation, we suggest interaction models for the~distribution of cubic crystal lattice orientations, where the~interaction is between pairs of orientations for neighbouring grains in the~tessellation. We discuss parameter estimation and model comparison methods based on maximum pseudolikelihood as well as graphical procedures for model checking using simulations.
Our methodology is applied for analysing a~dataset representing a~nickel-titanium shape memory alloy.  \\


\end{abstract}

\section{Introduction}

In a previous paper by some of us \cite{Seitl2022} we developed models for the~Laguerre tessellation describing the~morphology of polycrystalline datasets. The~present paper concerns statistical methodology for analysing crystallographic orientations (henceforth 'orientations') of grains conditional on that the~set of grains forms a~Laguerre tessellation. Including orientations as additional marks is crucial for applications in physics and material engineering, cf.\ \cite{Engler}, and leads us to consider complex interaction models for the~distribution of orientations \cite{Morawiec2004} conditioned on a~Laguerre tessellation, as opposed to previous studies based on models for the~distribution of individual orientations, see e.g.\ \cite{Arnold2018, Boogaart2002, Downs1972, Khatri1977, Leon2006}.

\subsection{Motivation and related results}

The microstructure of polycrystalline materials is characterized by the~morphology of grains and the~spatial redistribution of orientations among the~grains. 
The distribution of orientations
is often nonuniform and some preferential orientations occur. This phenomenon
is in material science called a~texture \cite{Engler}, and it determines the~elastic properties of the~material, which describe how much each grain deforms under external forces. 
The topic of fitting the~orientation distribution has a~long history, see e.g.\ \cite{Bohlke2006, Schaeben2017}. On the~one hand, nonparametric kernel estimation methods have been developed \cite{Boogaart2007}, which may be suitable 
for electron backscatter diffraction (3D-EBSD) measurements \cite{Zaefferer2009}. 
On the~other hand, parametric models and estimation of the~orientation probability density have been suggested \cite{Arnold2018}, and also maximum likelihood estimation for the~Bingham distribution using the~EM algorithm has been proposed \cite{Niezgoda2016}.

The deformation of grains is affected not only by the~orientation of each grain but also by the~orientations of its neighbouring grains \cite{Heller2020}. Since the~grains are 
space-filling with pairwise disjoint interiors,
they constitute a~tessellation, and a~very flexible class of tessellation models are so-called Laguerre tessellations \cite{Lautensack2008}. Statistical models and estimation procedures for Laguerre tessellations describing the~grain structure in polycrystalline materials have been developed in \cite{Seitl2022}. Moreover, using the~two statistical tests introduced in \cite{Pawlas2020}, it was shown that the~tessellation and orientations should not be generally treated separately. 
Nevertheless, in the~earlier work mentioned above, the~problem of the~estimation of the~orientation distribution is considered independently of the~tessellation, hence not taking into account the~geometry of the~grain structure.

Based on experimentally measured 3D microstructure obtained by X-ray diffraction (3D-XRD) \cite{Poulsen2004} or by 3D-EBSD \cite{Stanek20, Zaefferer2009}, physicists and material engineers study the~effect of the~orientation distribution on the~elastic properties of the~material \cite{Ding2020, Karafiatova2022}. However, as these are expensive methods of microscopy, we are motivated to propose a~complementary, less expensive statistical approach as developed in the~following.

\subsection{Our contribution and outline}
\label{sec:outline}

Our goal is to offer statistical models and simulation procedures for the~joint distribution of the~orientations when we condition on the~underlying tessellation. In particular, we account for the~dependencies between orientations of neighbouring grains. To the~best of our knowledge, such models have not been introduced before in the~literature. Due to the~complexity of the~data it may be too ambitious to aim at a~model fitting all aspects seen in the~data. 
Moreover, orientation characteristics as considered in this paper are of interest in material science research \cite{Ding2020, Karafiatova2022, Liu2014}. Therefore, when fitting models we will concentrate on these orientation characteristics as illustrated at the~end of this paper.

Perhaps our most important contribution is the~modelling part. We start with a~rather flexible semiparametric model not accounting for the~dependence on the~tessellation and with independence between grain orientations. Then we use the~Laguerre tessellation constructed
in \cite{Petrich2019}, and extend the~model to more complex models incorporating dependencies between the~tessellation and orientations: conditional on the~tessellation,  
we model dependencies between orientations for neighbouring grains; then we extend this by adding various weights as detailed in Section~\ref{sec:models}.

The paper is organized as follows. Section~\ref{sec:ori_repr} provides the~needed background for various orientation representations, where invariance under the~group of symmetries of a~cube is a~natural assumption, as well as orientation characteristics used for model specification and comparison.
Section~\ref{sec:data} introduces a~dataset of a~microstructure of metallic material used in later sections to illustrate our statistical approach. Section~\ref{sec:models} presents our three parametric classes, followed in Section~\ref{sec:methods} by methods for parameter estimation, model comparison and model checking as well as simulation. The~methodology in Sections~\ref{sec:models} and \ref{sec:methods} is illustrated in Section~\ref{sec:results}. 

\section{Orientation representations and characteristics}
\label{sec:ori_repr}

\subsection{The fundamental region for the~case of a~cubic lattice structure}
\label{sec:fundamental}

The kind of data we study in this paper are `crystallographic orientations' observed in polycrystalline materials where each grain is comprised of atoms forming a~cubic lattice structure \cite{Engler}. To make the~meaning of this more precise we need to introduce an~equivalence relation and some notation. Let $C_1,\dots,C_n$ be a~finite collection of grains which are 
closed 3-dimensional sets, their interiors are pairwise disjoint and their union $W=C_1\cup\ldots\cup C_n$ is a~connected set in 3-dimensional Euclidean space, which we equip with a~coordinate system with axes $x$, $y$ and $z$. In other words, the~grains are the~cells of a~tessellation (subdivision) of $W$ which we refer to as the~`specimen'. 
Let $\mathrm{SO}(3)$ be the~special orthogonal group, that is,  the~set of $3\times3$ rotation matrices with determinant~$1$. Let $\mathcal{O}$ be the~subgroup of $\mathrm{SO}(3)$ 
given by the~symmetries of a~cube. There are 48 symmetries and $\mathcal{O}$ has 24 elements \cite{Mueller2013}. We consider two elements $G_1$ and $G_2$ of $\mathrm{SO}(3)$ to be equivalent if $G_1=HG_2$ for some $H\in\mathcal O$. The~equivalence classes of this relation 
form a~partition of $\mathrm{SO}(3)$ and 
a transversal is a~set of 
 
representatives in the~equivalence class sense. Below we define a~particular set $F$ 
which we call the~fundamental zone and which 'effectively' is a~transversal (more precisely, apart from a~nullset with respect to the~measure in \eqref{eq:Haar} below, $F$ is a~transversal whose more complicated definition is given in \ref{sec:transversal}). 
By a~crystallographic orientation we mean an~element of $\mathrm{SO}(3)$ or just of $F$ because of the~equivalence relation.

It is convenient to represent every matrix $G\in\mathrm{SO}(3)$ by its Euler angles $\left( \varphi_1,\, \phi,\, \varphi_2 \right) \in [0,2\pi) \times [0,\pi] \times [0,2\pi)$ (determining sequential rotations around the~$z$-axis, the~$x$-axis and again the~$z$-axis, see \cite[Section 2.6]{Engler}). Henceforth, we identify $G$ and $\left( \varphi_1,\, \phi,\, \varphi_2 \right)$ without any mentioning, and we use the~symbol $g$ as a~general notation for both. 

Now, define the~fundamental zone by
\begin{equation}
\label{eq:fz}
F = \big\{
(\varphi_1,\,\phi,\,\varphi_2) \mid
\varphi_1 \in [0,\, 2\pi),
\ \varphi_2 \in [0,\,\pi/2), 
\ \phi \in [ \phi_0 (\varphi_2),\, \pi/2] 
\big\},
\end{equation}
where
\[
\phi_0 (\varphi_2)  = \arccos ~\text{min} 
\left( \dfrac{\cos \varphi_2}{\sqrt{\strut 1+ \cos^2 \varphi_2}}, \dfrac{\sin \varphi_2}{\sqrt{\strut 1+ \sin^2 \varphi_2}} \right).
\]

\subsection{Random orientation representations}
\label{sec:quotient}

By a~random orientation we understand a~random variable with values in the~corresponding space, that is, either $\mathrm{SO}(3)$ for an~orientation matrix representation or $[0,2\pi) \times [0,\pi] \times [0,2\pi)$ for an~Euler angles representation, where these spaces are equipped with the~corresponding Borel $\sigma$-algebra. Since we pay attention to crystallographic orientation in the~setting of Section~\ref{sec:fundamental}, we restrict attention to distributions on $F$. Then, by adding an~independent uniform distribution on $\mathcal O$, this easily extends to a~distribution on $\mathrm{SO}(3)$. 
 
For notional convenience we do not distinguish between whether $G, \varphi_1,\phi,\varphi_2$ or another variable introduced in the~following refers to a~realisation of some random variable, the~random variable itself or an~argument of a~function describing some properties of the~random variable. Its meaning will always be obvious from the~context.

Let $\mu$ be the~normalized (that is, $\mu$ is a~probability measure) Haar measure \cite{Haar1933} on $\mathrm{SO}(3)$. Then the~random orientation distributed according to $\mu$ has the~uniform distribution and in terms of Euler angles, for any Borel set $B\subseteq [0,2\pi) \times [0,\pi] \times [0,2\pi)$,
\begin{equation}\label{eq:Haar}
\mu(B) 
= \frac{1}{8 \pi^2} \int_0^{2 \pi} \int_0^{\pi} \int_0^{2 \pi} \sin(\phi) \bm{1}{\left[ (\varphi_1,\phi,\varphi_2) \in B \right]} \, {\rm d} \varphi_1 \, {\rm d}\phi \, {\rm d}\varphi_2,
\end{equation}
where $\bm{1}[\cdot]$ is the~indicator function.
Thus the~corresponding random Euler angles $\varphi_1, \phi, \varphi_2$ are independent, with $\varphi_1$ and $\varphi_2$ uniformly distributed on $[0,2\pi)$, and $\eta = \cos\phi$ uniformly distributed on $[-1,1]$. 

In later sections we consider various probability density functions. When a~single orientation is considered, the~density is always with respect to $\mu_F$, the~restriction of $\mu$ to $F$. When a~vector of $n$ orientations is considered, the~density is always with respect to the~product measure of $\mu_F$ $n$ times.

\begin{figure}[tb]
\centering
\begin{minipage}[t]{0.49\textwidth}
\centering
\includegraphics[width=\textwidth]{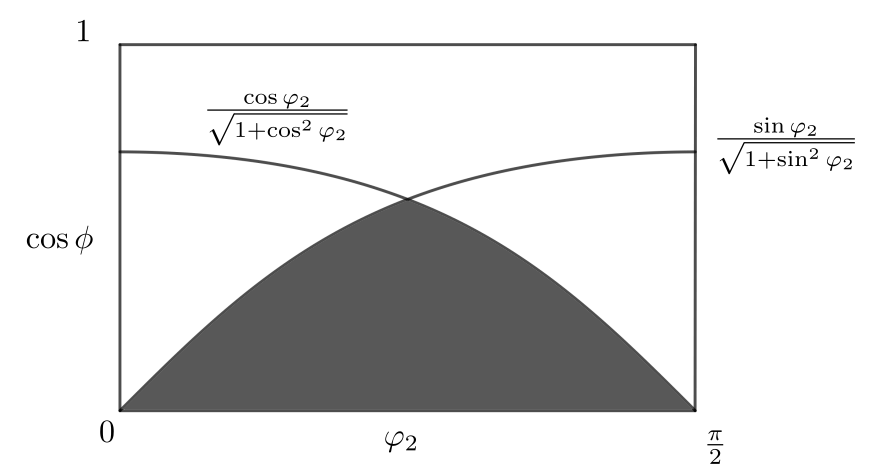}
\caption{A cross-section of $t(F)$ (black area) given by \eqref{eq:transformed_fz} for an~arbitrary $\varphi_1\in [0,2\pi)$.}
\label{fig:transformed_fz}
\end{minipage}
\hfill
\begin{minipage}[t]{0.49\textwidth}
\centering
\includegraphics[width=0.6\textwidth]{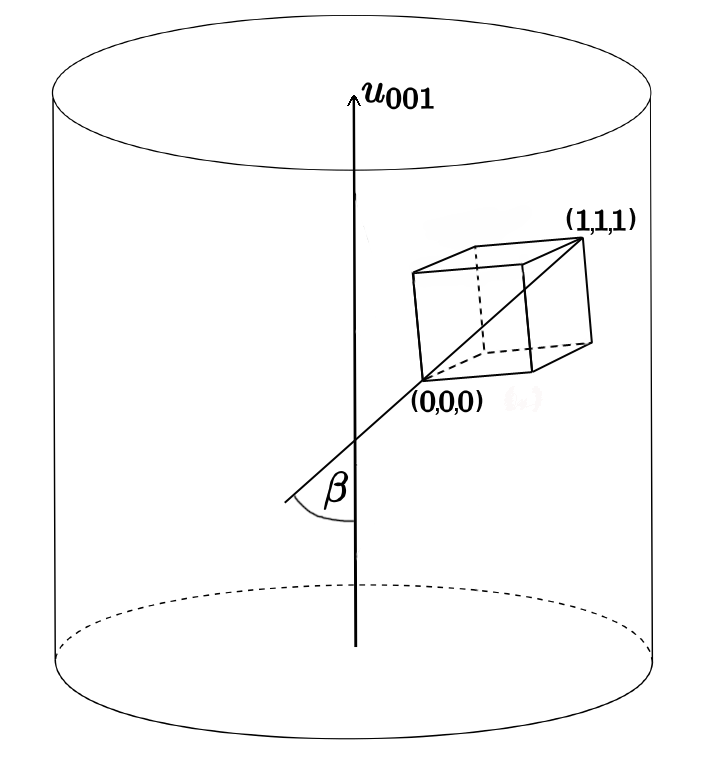}
\caption{The angle $\beta$ specifies the~tilt given by $\cos \beta$ of $v=(1,1,1)^\top$ with respect to a~particular orientation $G$, cf. \eqref{eq:tilt}. Here, the~specimen is assumed to be a~cylinder with a~coordinate system such that the~$z$-axis agrees with the~axis of the~cylinder.}
\label{fig:tilt}
\end{minipage}
\end{figure}

In many places of this paper it becomes  useful to consider the~`transformed orientation' $t(\varphi_1,\phi,\,\varphi_2)=(\varphi_1,\, \eta,\,\varphi_2)$ living in the~`transformed fundamental zone' given by
\begin{equation}\label{eq:transformed_fz}
 t(F) = \big\{
(\varphi_1,\, \eta,\,\varphi_2) \mid
\varphi_1 \in [0,\, 2\pi),
\ \varphi_2 \in [0,\,\pi/2), 
\ \eta \in \left[ 0, \cos \phi_0 (\varphi_2) \right] 
\big\}.   
\end{equation}
It follows from \eqref{eq:Haar} that for any Borel set $B\subseteq t(F)$ the~measure $\mu_F$ transformed by $t$ is given by
\begin{align}
\label{eq:Leb}
\begin{split}
t\mu_F(B) & =  \int_{t(F)} \bm{1}{\left[ (\varphi_1,\eta,\varphi_2) \in B \right]}  \, {\rm d}(\varphi_1, \eta, \varphi_2) \\
& = \int_0^{2 \pi} \int_0^{\pi/2} \int_0^{\cos \phi_0 (\varphi_2)} \bm{1}{\left[ (\varphi_1,\eta,\varphi_2) \in B \right]}  \, {\rm d}\eta \, {\rm d} \varphi_2 \, {\rm d}\varphi_1.
\end{split}
\end{align}
Figure~\ref{fig:transformed_fz} shows a~graphical representation of $t(F)$ by a~cross-section for an~arbitrarily chosen value of $\varphi_1$. 

\subsection{Orientation characteristics}
\label{sec:chars}

This section introduces various orientation characteristics where we start by considering a~single orientation, then a~pair of orientations, and finally a~sample of orientations.

To characterize an~orientation, some low-dimensional characteristics can be used, e.g.\ a~`tilt' 
\begin{align}
\label{eq:tilt}
\text{tilt}(G; v) = \max_{S \in\, \mathcal{O}}  \frac{v^\top S G u_{001}}{ \| v \|}, 
\end{align}
where $G$ is an~orientation matrix of a~grain, $v$ is a~chosen direction with respect to a~coordinate system specifying the~cubic lattice structure of the~grain and $u_{001} = (0,0,1)^\top$ corresponds to the~$z$-axis of the~specimen (this is the~choice for our application but any unit vector could be used instead).
In other words, in order to compute $\text{tilt}(G; v)$, we first compute the~cosine of angles between the~$z$-axis (in the~coordinate system for the~specimen) and all equivalent directions of $v$ according to the~orientation matrix $G$. Then we choose the~largest value. For the~data analyzed in this paper, as argued in Section~\ref{sec:data}, we choose $v$ to be the~main diagonal of the~cubic lattice, $v=(1,1,1)^\top$, see Figure~\ref{fig:tilt}.

A natural question is how to measure the~closeness of two orientations. 
The most common measure for two orientations $g_1$ and $g_2$ is the~`disorientation angle' \cite[Section 2.7.2]{Engler} which is given by
\begin{align}
\label{eq:dis_angle}
    \dis(g_1, g_2) = \min_{S \in \mathcal{O}} \arccos \frac{{\rm tr} \left(G_1^{-1} S G_2 \right) - 1}{2},
\end{align}
where $G_1$ and $G_2$ are the~orientation matrix representations corresponding to $g_1$ and $g_2$, respectively, and ${\rm tr}$ means the~trace of a~matrix.
Similar orientations have a~disorientation angle close to 0 and the~maximum disorientation angle is approximately $62.8^\circ$.
Incidentally, the~disorientation angle follows the~Mackenzie distribution if $g_1$ and $g_2$ are considered to be independent random variables following the~`uniform' distribution given by the~Haar measure in \eqref{eq:Haar} \cite[Section 7]{Morawiec2004}.

In \eqref{eq:dis_angle} all 24 elements of $\mathcal O$ must be considered. Another measure of closeness of two orientations, which is faster to compute, is the~inner product of the~embedding $\mathbf{t}$ introduced in \cite{Arnold2018} and given by
\begin{align}
\label{eq:inner_prod}
   \inn(g_1, g_2) = \langle \mathbf{t}(G_1), \mathbf{t}(G_2) \rangle = \sum\limits_{i=1}^3 \sum\limits_{j=1}^3 \left( G_{1i}^{\top} G_{2j} \right)^4 - \frac{9}{5},
\end{align}
where $G_{ki}$ is the~$i$th row of the~matrix $G_k$, $k=1,2$ (since the~definition of $\mathbf{t}$ is rather technical, we refer  to \cite{Arnold2018} for details). On the~contrary to the~disorientation angle, this inner product can be both negative and positive, and closeness corresponds to high values (the maximum value is $6/5$). 

\begin{figure}[tb]
\centering
\begin{subfigure}{0.45\textwidth}
\centering
\includegraphics[width=\textwidth]{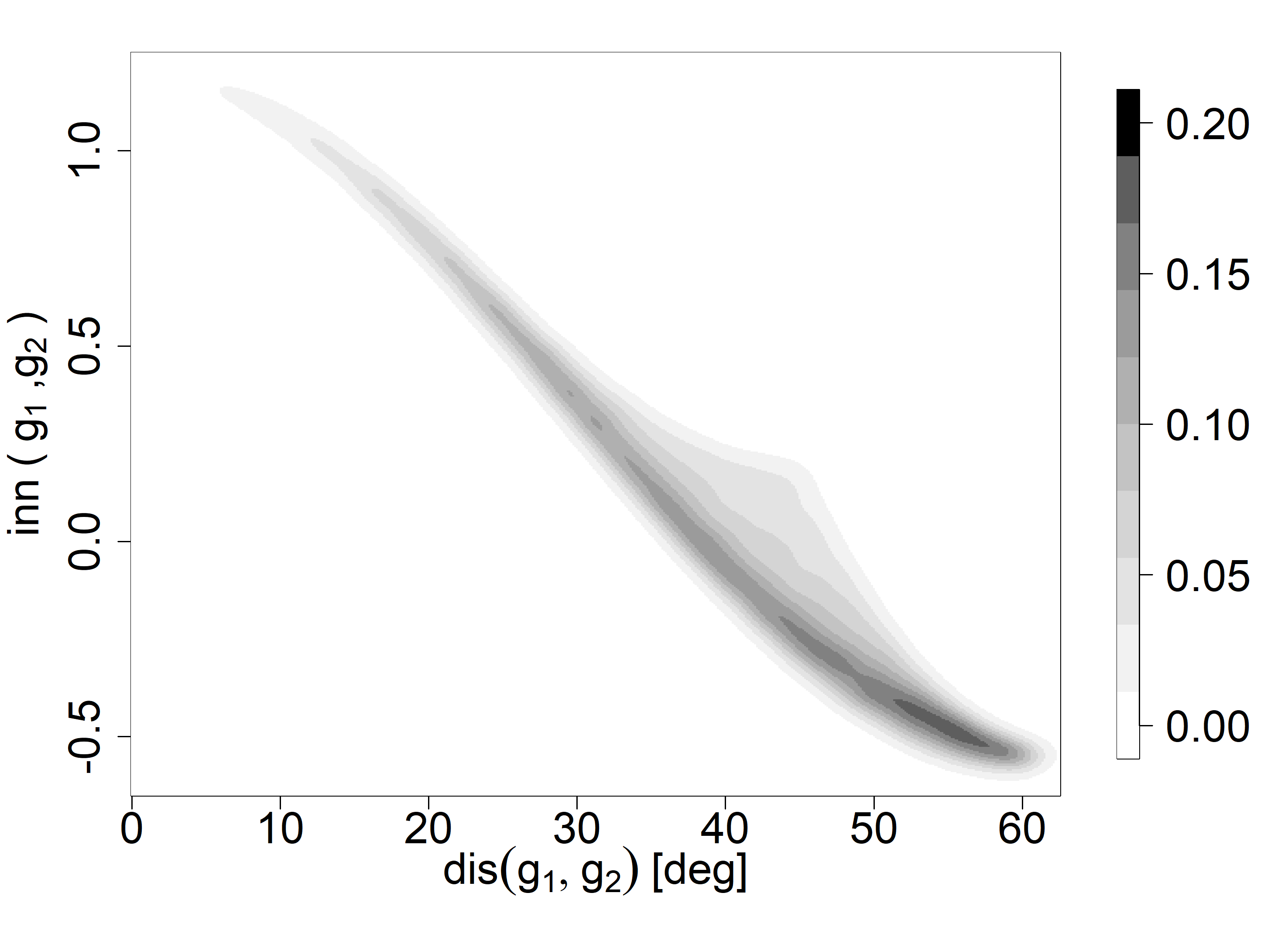}
\caption{Uniform distribution.}
\label{fig:dis_inn_uni}
\end{subfigure}
\begin{subfigure}{0.45\textwidth}
\centering
\includegraphics[width=\textwidth]{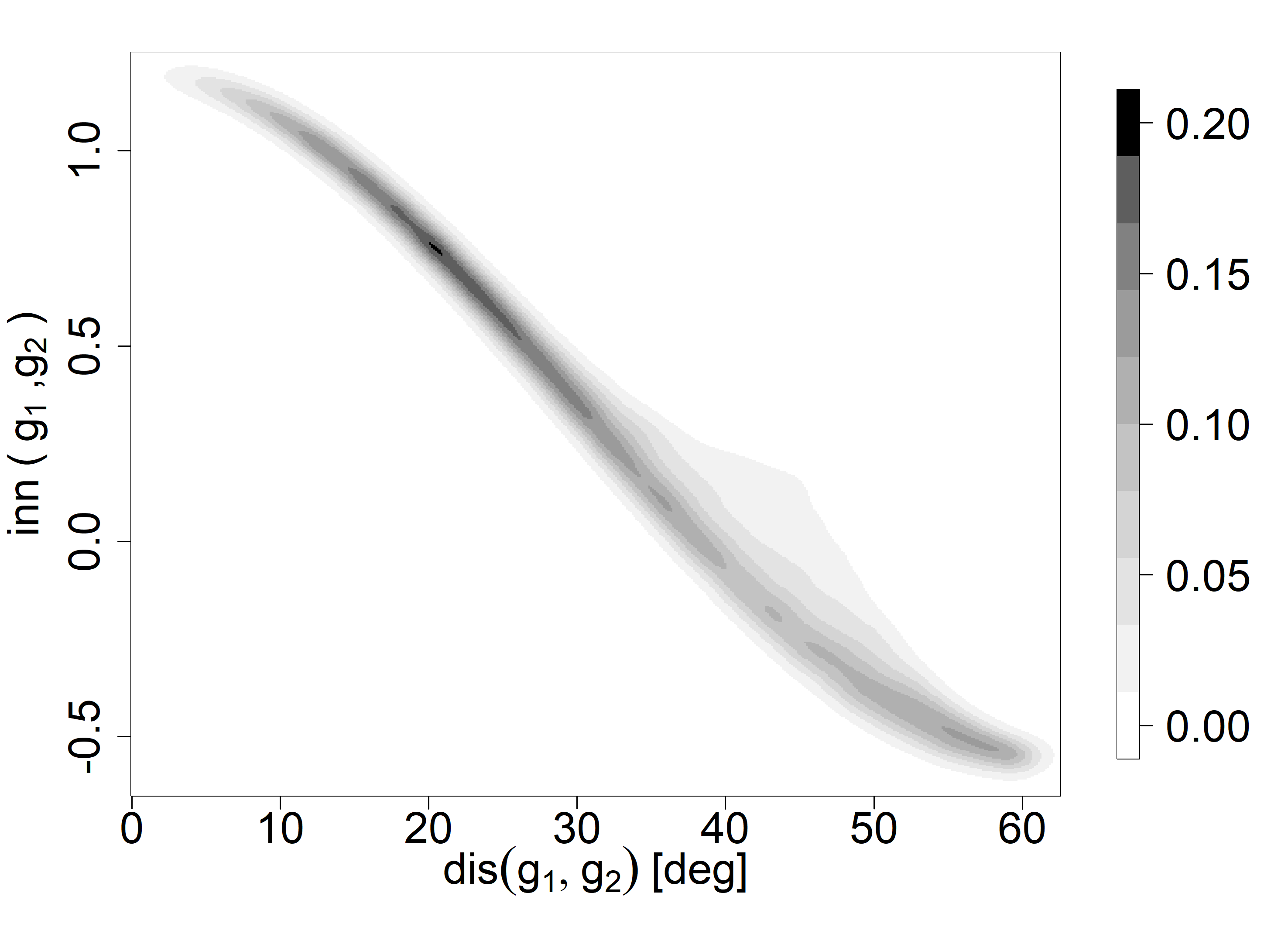}
\caption{NiTi data.}
\label{fig:dis_inn_NiTi}
\end{subfigure}
\caption{Kernel density estimators of the~bivariate distribution of the~inner product $\inn(g_1,g_2)$ and the~disorientation angle $\dis(g_1,g_2)$ from (a) the~simulated uniform orientation distribution and (b) the~NiTi dataset (introduced in Section~\ref{sec:data}).}
\end{figure}

To understand the~relationship between the~measures in \eqref{eq:dis_angle} and \eqref{eq:inner_prod}, we simulated their joint distribution when $g_1$ and $g_2$ are considered to be independent random variables following
the~Haar measure in \eqref{eq:Haar}, see Figure~\ref{fig:dis_inn_uni}. 
As expected the~disorientation angle and the~inner product are strongly negatively correlated, with the~Spearman rank correlation coefficient equal to $-0.969$. 

Finally, for a sample $g_1,\dots,g_n$ of orientations, \cite{Arnold2018} introduced a~certain scalar $\hat{d}$ which has an~interpretation as 
a sample dispersion (its precise definition is technical, so again we refer to \cite{Arnold2018}). For later use, note that $0\le \hat{d}\le 6/5$, where $\hat{d}=0$ if the~$g_i$ are equal.

\section{Data example}
\label{sec:data}

In this section we introduce the~data used to illustrate our methodology in the~rest of the~paper. Briefly, the~data comes from \cite{Sedmak2016} and concerns an~experiment of a~nickel-titanium (NiTi) wire evaluated by the~{3D-XRD} method. The~centroids, volumes and orientations were determined for the~microstructure, but no information about the~neighbouring structure was provided, so 
we obtained this information using the~construction of the~Laguerre tessellation in the~article \cite{Petrich2019}.
The~data presents a~cutout consisting of $n=1060$ grains, the~centroids of which lie in a~rectangular parallelepiped of size $44 \times 44 \times 40 \, \mu {\rm m}^3$, situated in the~central part of the~wire. Henceforth, we use the~notation $\C_n=(C_1,\dots,C_n)$ for the~grains (or cells) of the~tessellation and $\g_n=(g_1,\dots,g_n)$ for the~corresponding orientations. 

In \cite{Heller2020} it was argued using graphical methods that the~orientations are not uniformly distributed.
In particular, the~authors noticed that the~orientations are concentrated so that the~$z$-axis of the~specimen is aligned with a~crystal direction given by the~main diagonal of a~cube (direction $(1,1,1)^\top$). This alignment motivates us to use the~tilt in \eqref{eq:tilt} with ${v=(1,1,1)^\top}$ as a~characteristic when we consider a~model checking procedure in Section~\ref{sec:results}. 

\begin{figure}[tb]
\centering
\begin{subfigure}{0.3\textwidth}
\includegraphics[width=\textwidth]{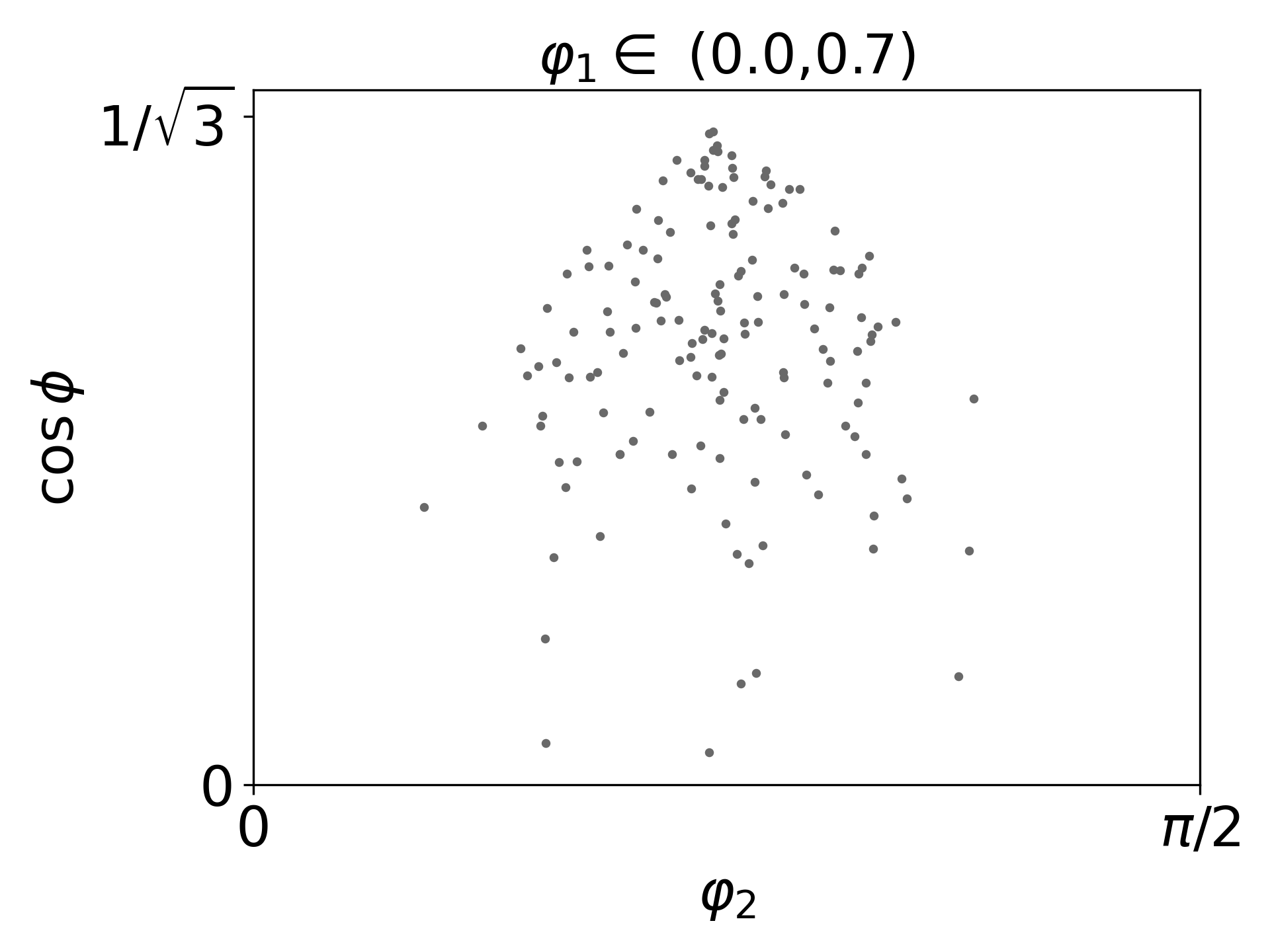}
\end{subfigure}
\begin{subfigure}{0.3\textwidth}
\includegraphics[width=\textwidth]{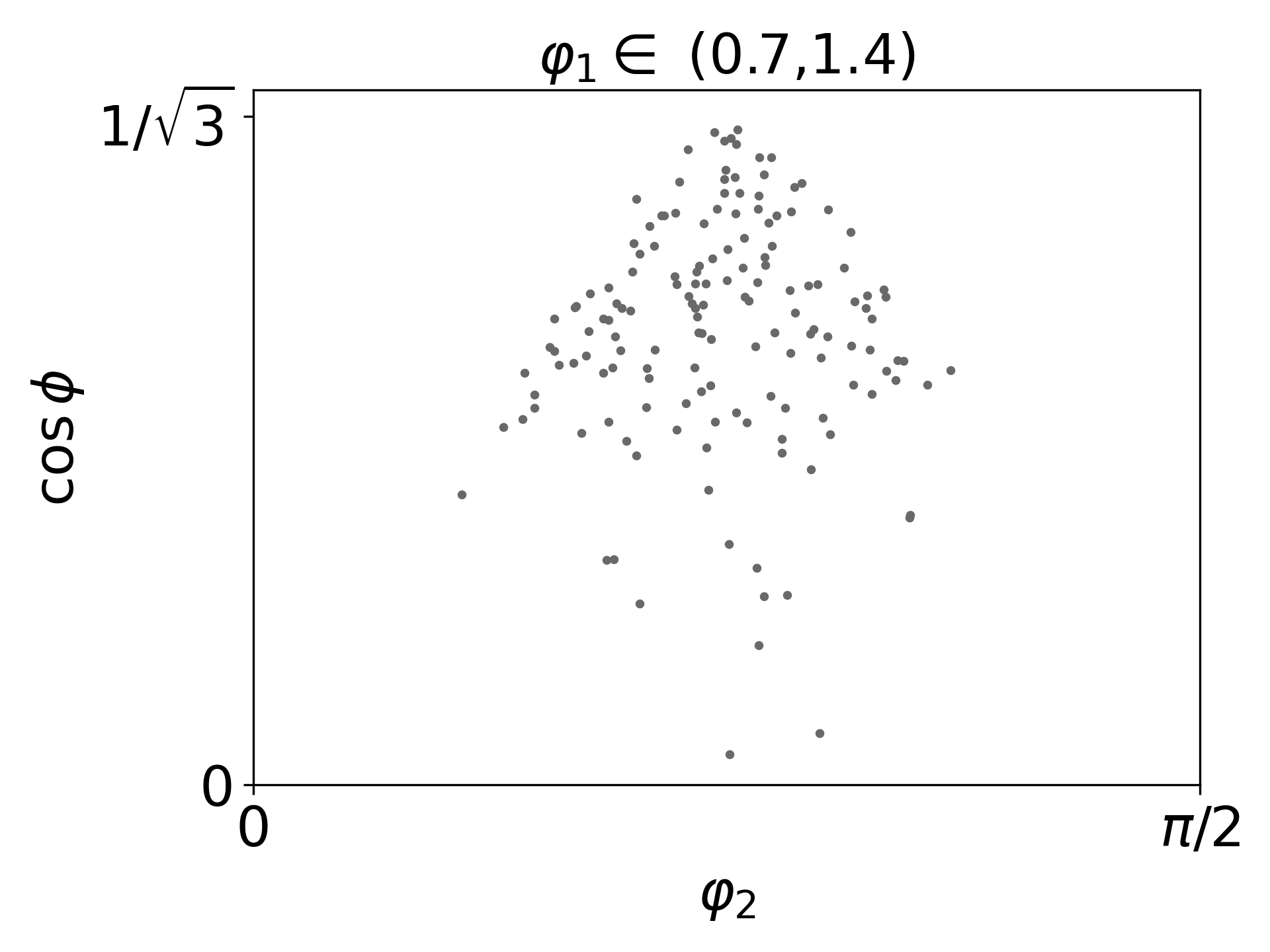}
\end{subfigure}
\begin{subfigure}{0.3\textwidth}
\includegraphics[width=\textwidth]{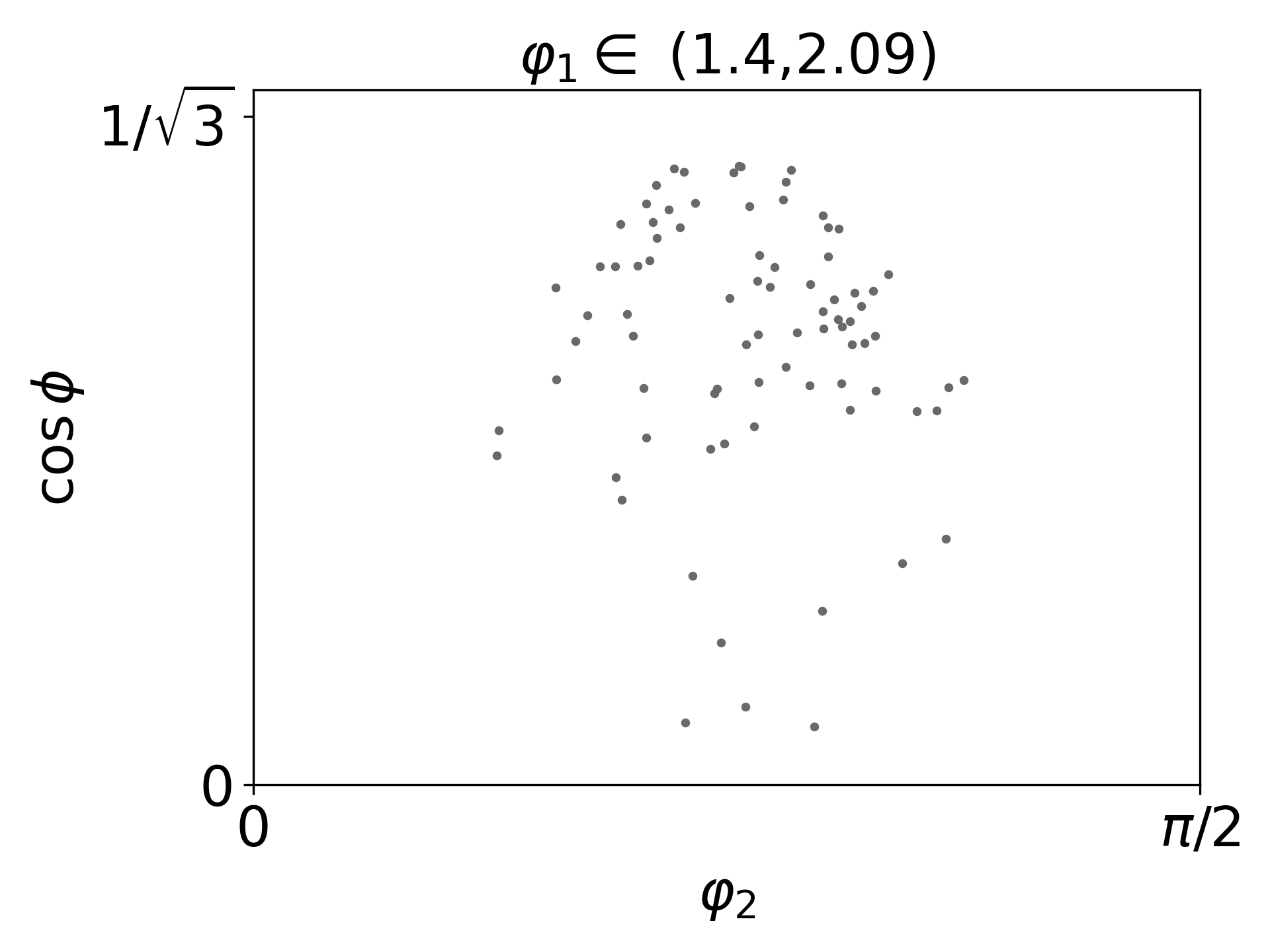}
\end{subfigure}
\begin{subfigure}{0.3\textwidth}
\includegraphics[width=\textwidth]{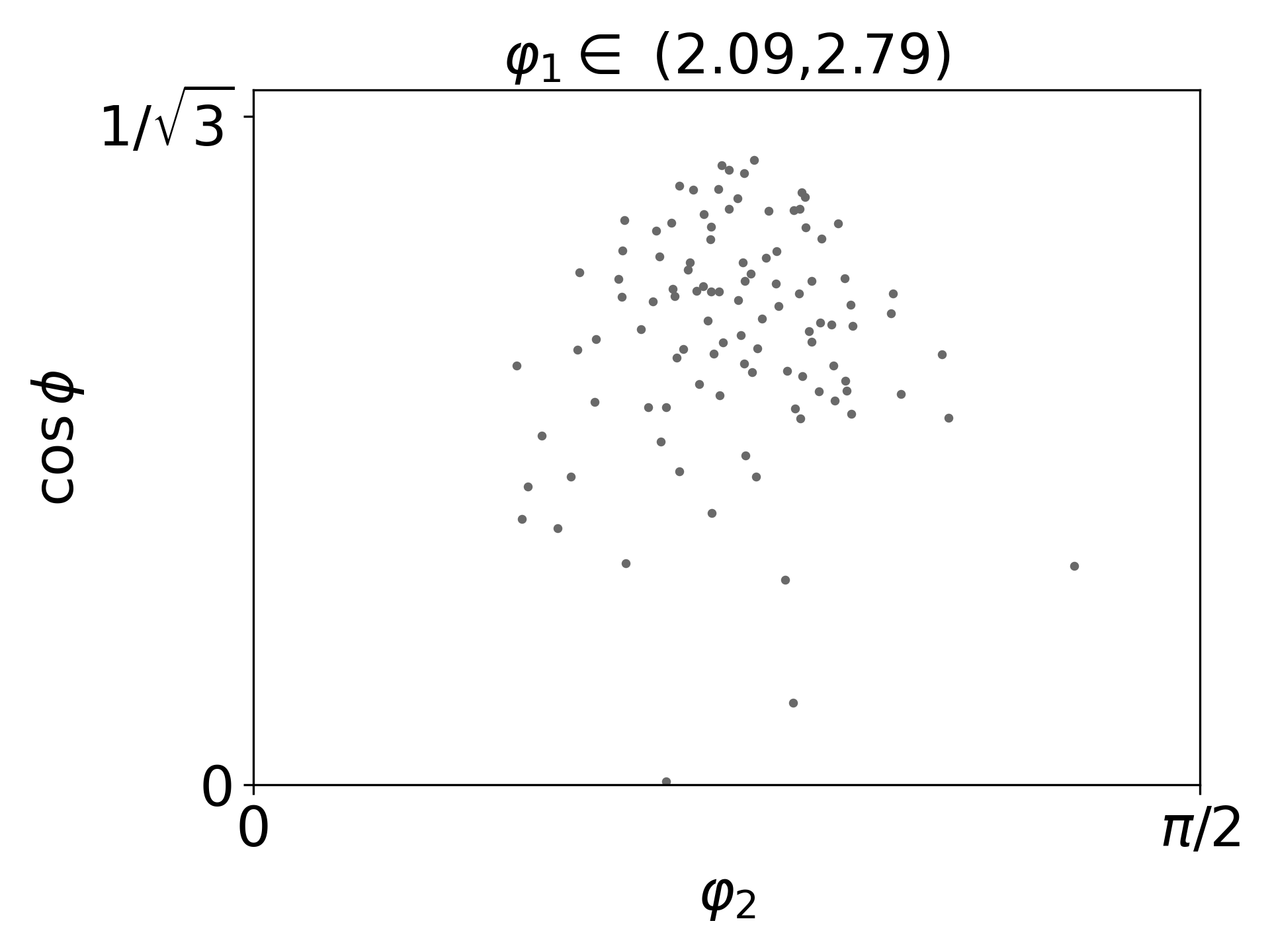}
\end{subfigure}
\begin{subfigure}{0.3\textwidth}
\includegraphics[width=\textwidth]{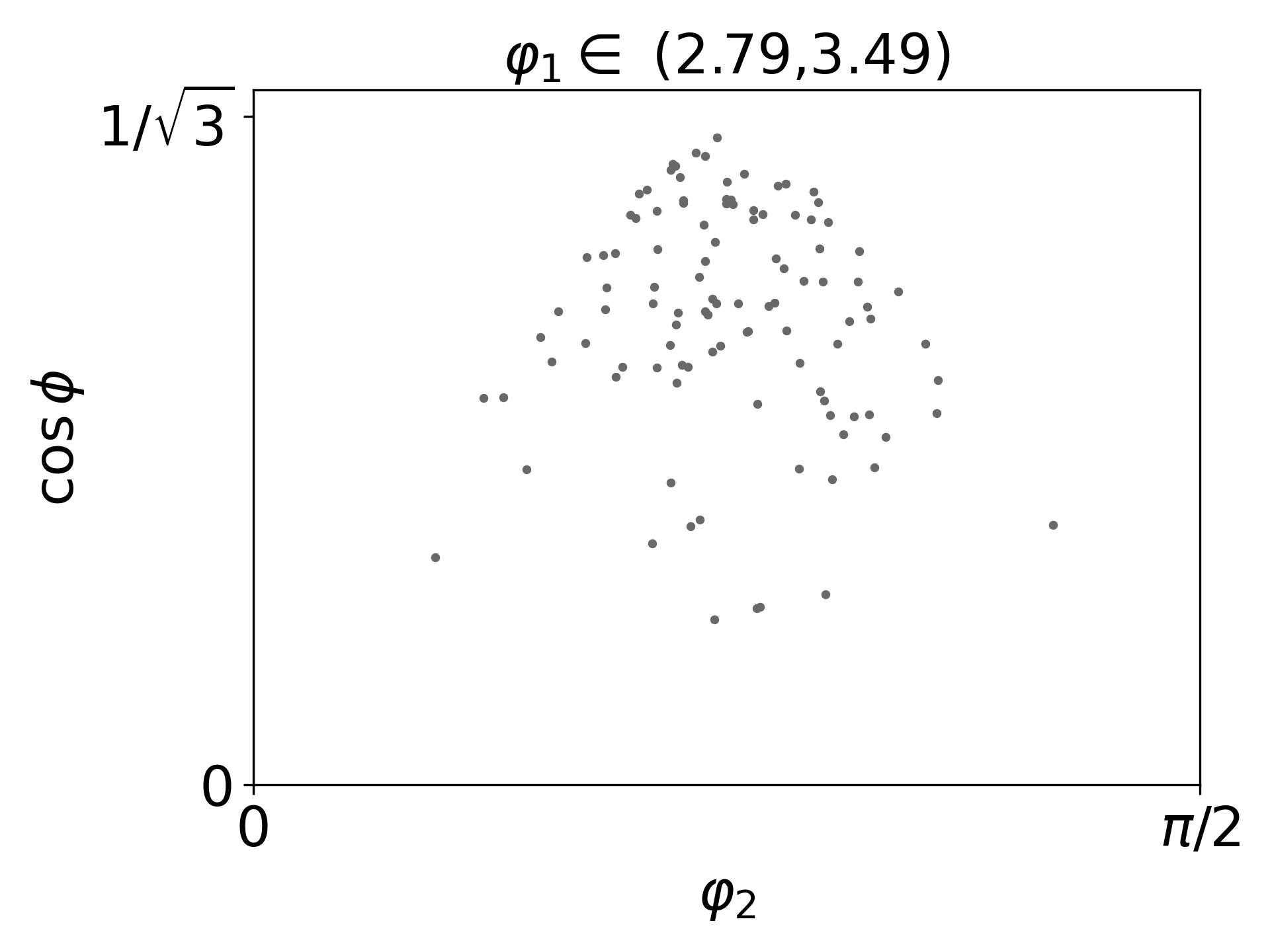}
\end{subfigure}
\begin{subfigure}{0.3\textwidth}
\includegraphics[width=\textwidth]{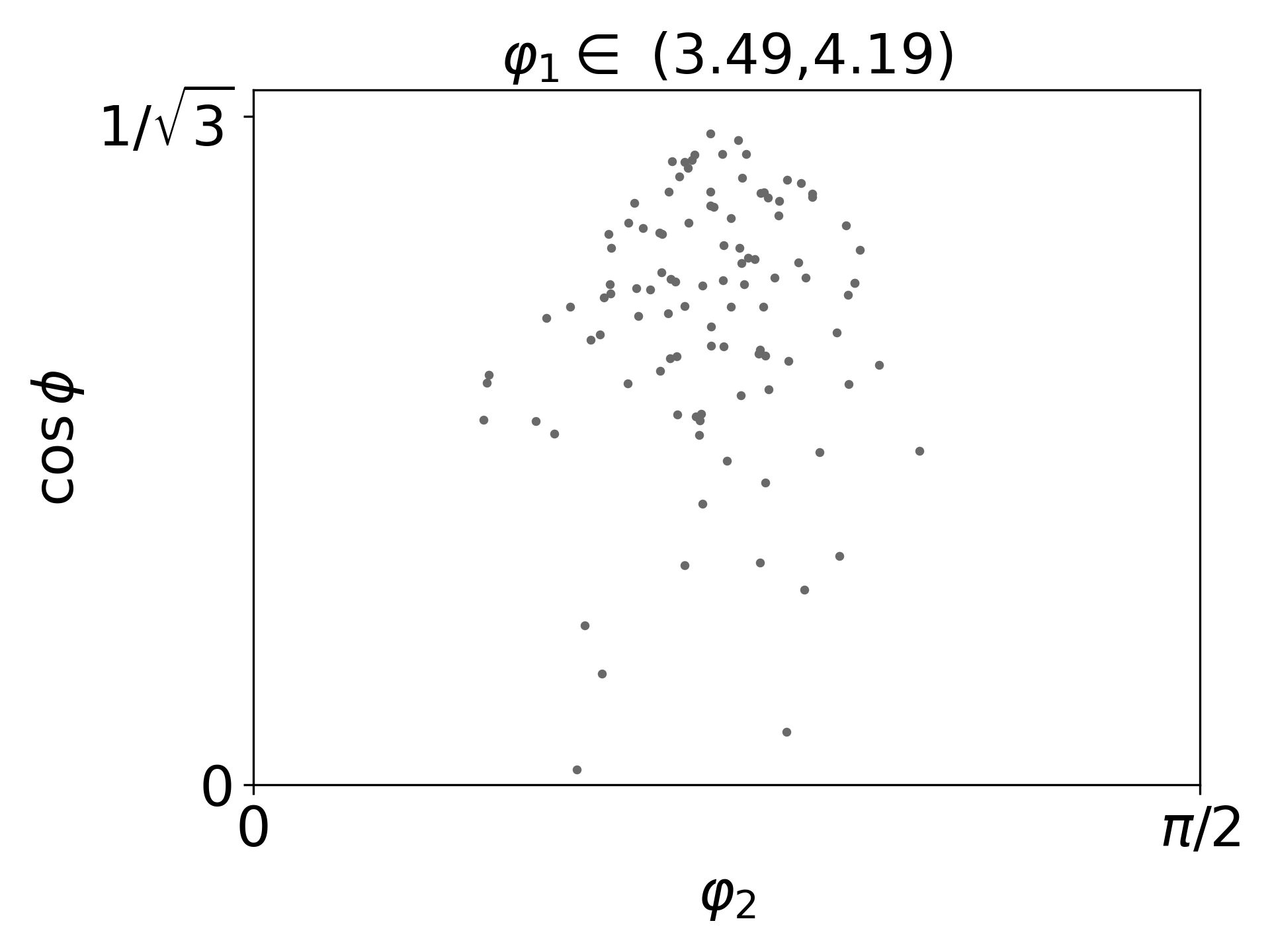}
\end{subfigure}
\begin{subfigure}{0.3\textwidth}
\includegraphics[width=\textwidth]{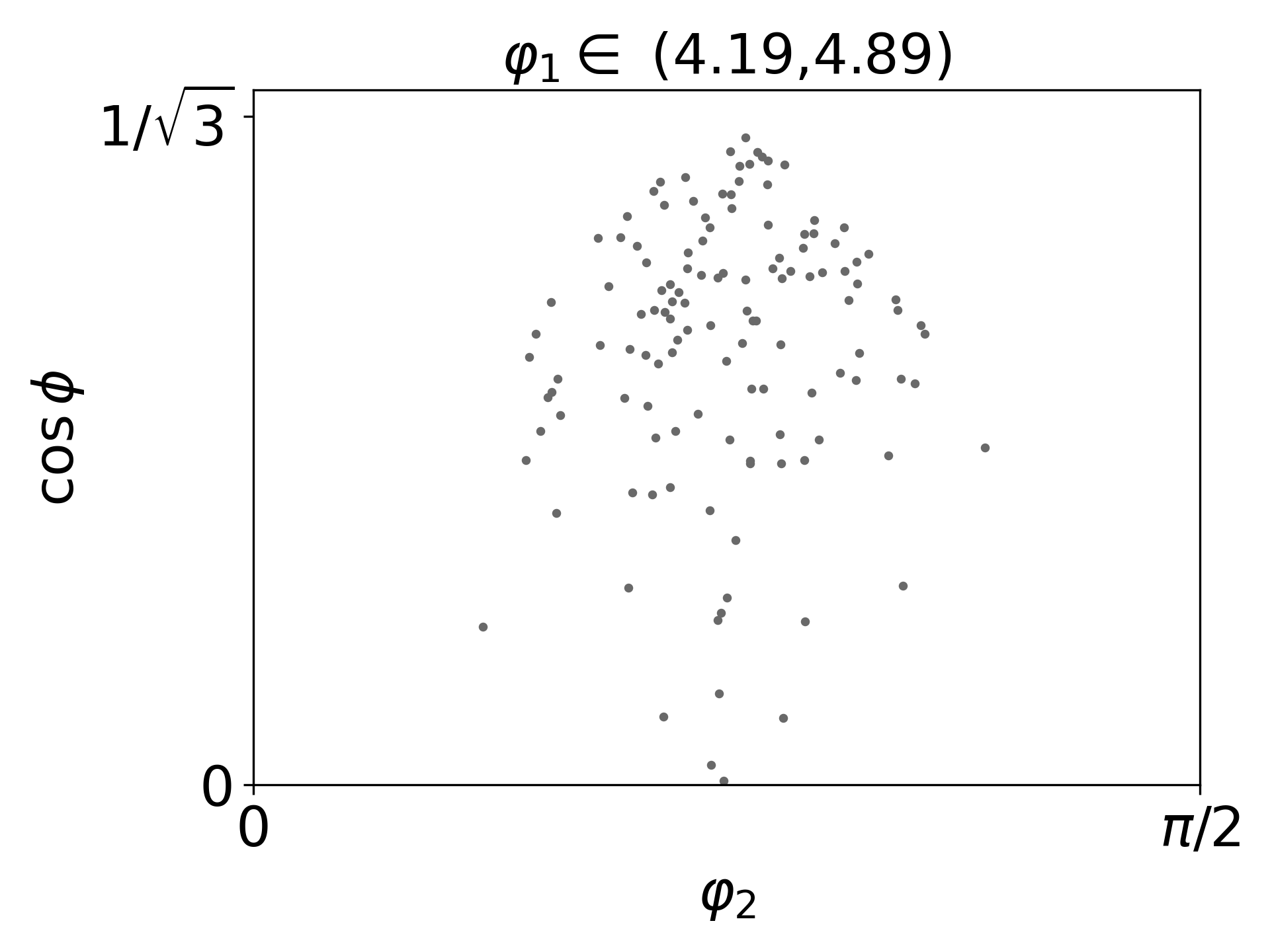}
\end{subfigure}
\begin{subfigure}{0.3\textwidth}
\includegraphics[width=\textwidth]{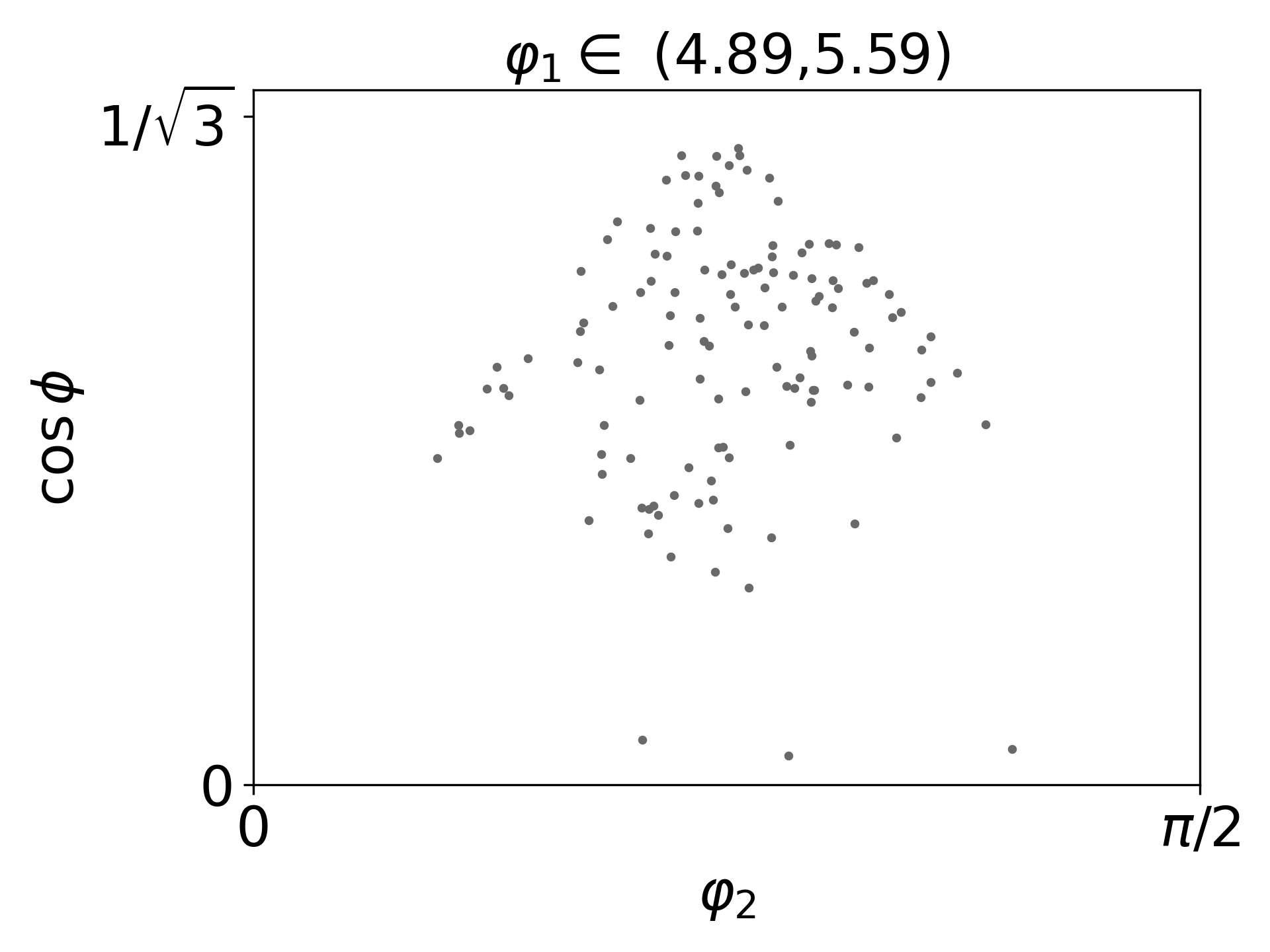}
\end{subfigure}
\begin{subfigure}{0.3\textwidth}
\includegraphics[width=\textwidth]{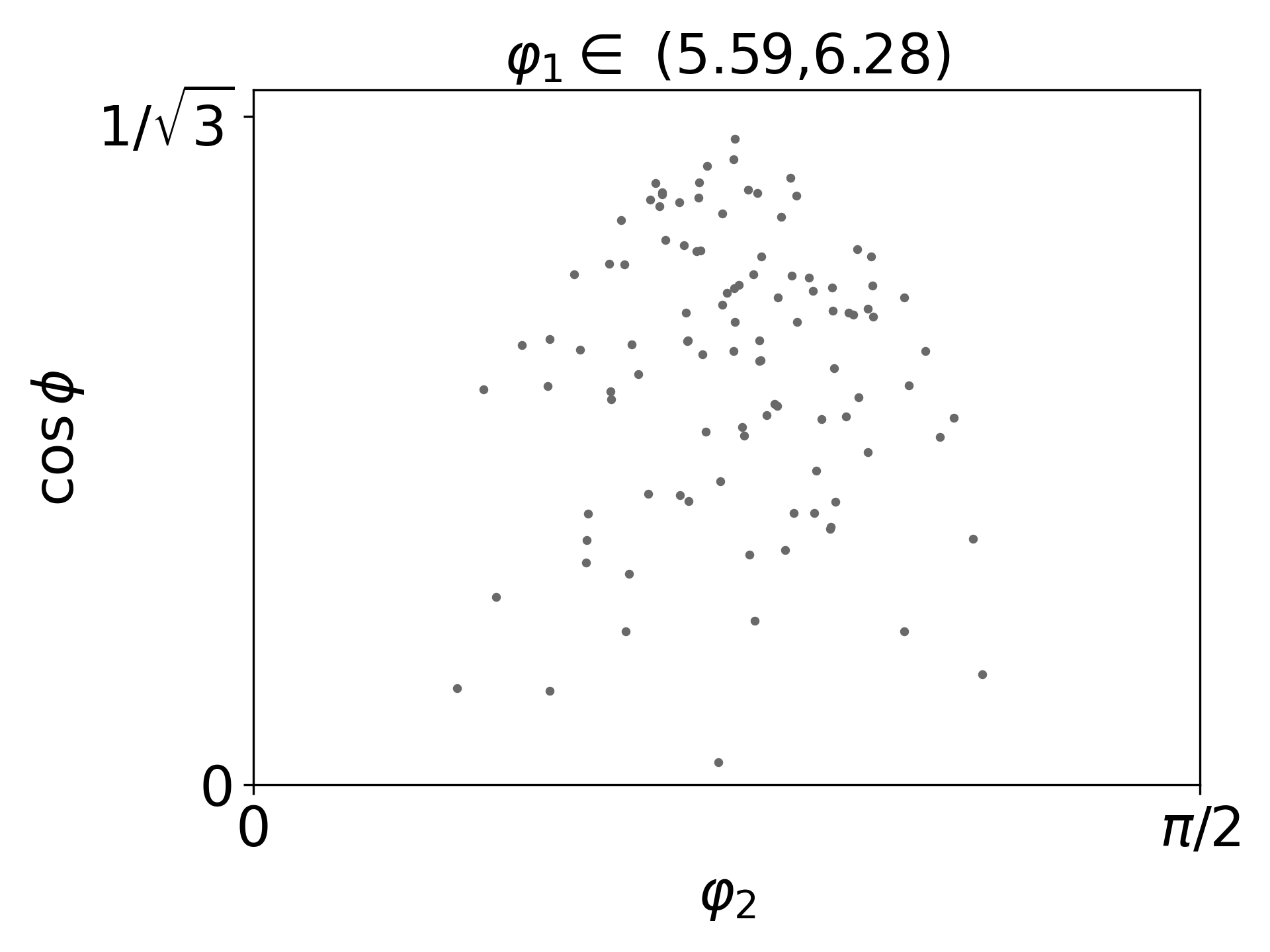}
\end{subfigure}
\caption{Transformed Euler angles for the~NiTi dataset: Scatter plots of pairs $(\cos\phi, \varphi_2)$ with the~corresponding $\varphi_1$ in one of nine intervals as stated on the~top of each plot.}
\label{fig:cross_sections_NiTi}
\end{figure}

To visualise the~transformed orientations within the~transformed fundamental zone, we first divide the~domain of $\varphi_1$ into $9$ subintervals $[a_i,a_{i+1})$ of equal length, so ${a_i=2\pi i/9}$, $i=0,\dots,9$. For each $i=0,\dots,8$ we obtain a~scatter plot of pairs $(\cos \phi,\, \varphi_2)$ with the~corresponding $\varphi_1$ in $[a_i,a_{i+1})$. Figure~\ref{fig:cross_sections_NiTi} shows such cross-sections for the~data in hand.
We observe a~higher concentration of points around the~larger values of $\cos \phi$. This indicates the~presence of a~preferential orientation (but it is implausible to determine the~specific preferential orientation from the~figure).

Figure~\ref{fig:dis_inn_NiTi} shows the~relationship between the~inner product \eqref{eq:inner_prod} and the~disorientation angle \eqref{eq:dis_angle} for the~data.
In comparison to Figure~\ref{fig:dis_inn_uni}, a~slightly stronger negative correlation takes place as the~Spearman rank correlation coefficient is $-0.991$ and the~disorientation distribution is shifted to the~left.

\section{Models}
\label{sec:models}

In continuation of the~model of random Laguerre tessellation distribution from \cite{Seitl2022}, we now look for a~suitable model for the~conditional distribution of the~orientations $\g_n$ given the~Laguerre tessellation $\C_n$. We start in Section~\ref{sec:simple} with a~semiparametric model for the~Euler angles, where its probability density function (pdf) is fully specified. Then we expand on this model in Section~\ref{sec:int} by introducing three parametric models which account for the~interaction between orientations of neighbouring grains. Estimation of the~parameters of these three models is deferred to Section~\ref{sec:methods}.

\subsection{A semiparametric model for the~Euler angles}
\label{sec:simple}

In this section we assume for simplicity that $\g_n$ is independent of $\C_n$ and $g_1,\dots,g_n$ are independent and identically distributed. Hence we are left with specifying a~model for a~single grain with Euler angles $(\varphi_1,\phi,\varphi_2)$ within the~fundamental zone $F$.

\begin{figure}[tb]
\centering
\begin{subfigure}{0.45\textwidth}
\centering
\includegraphics[width=0.7\textwidth]{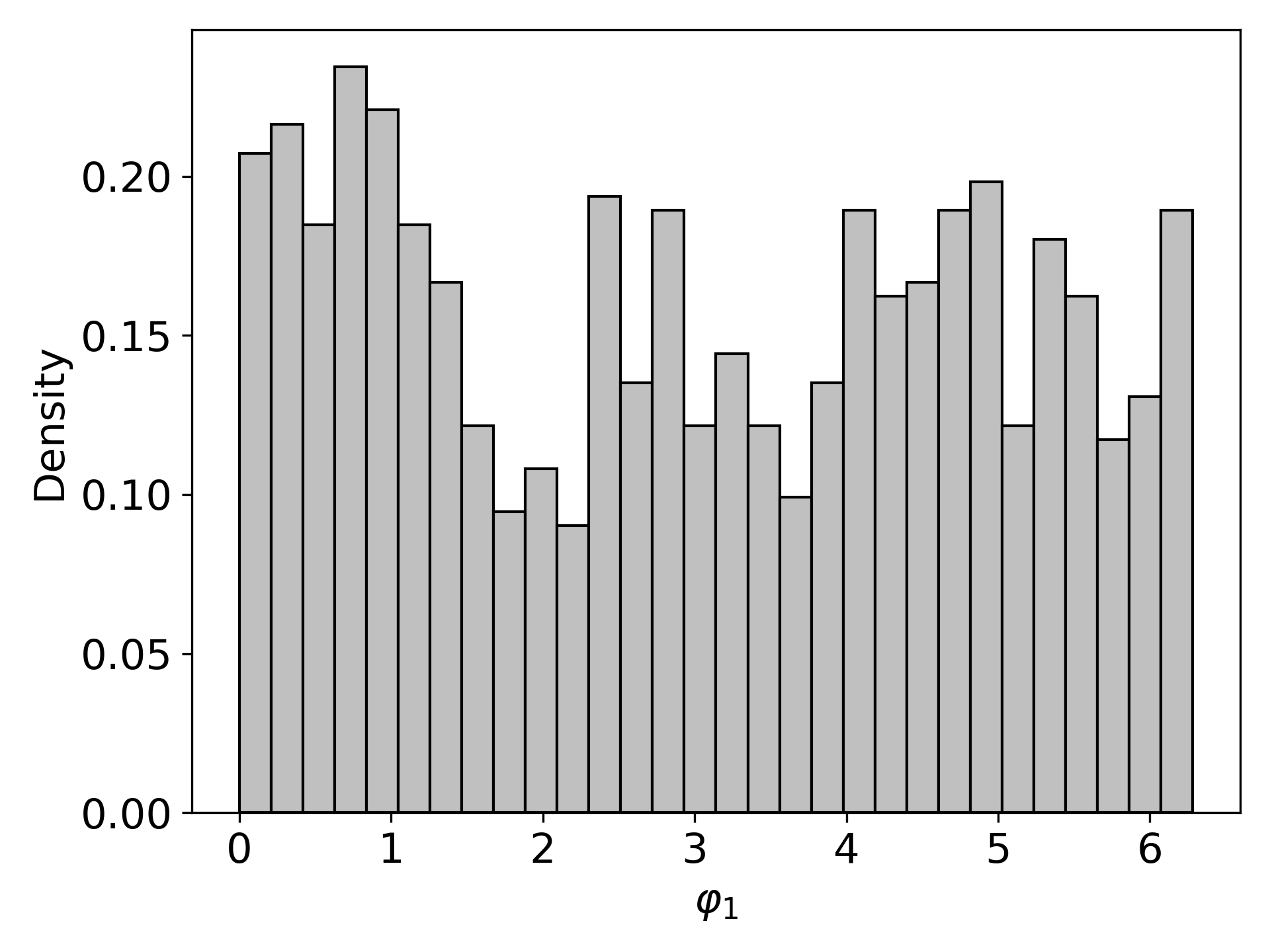}
\caption{}
\label{fig:hist_phi1}
\end{subfigure}
\begin{subfigure}{0.45\textwidth}
\centering
\includegraphics[width=0.7\textwidth]{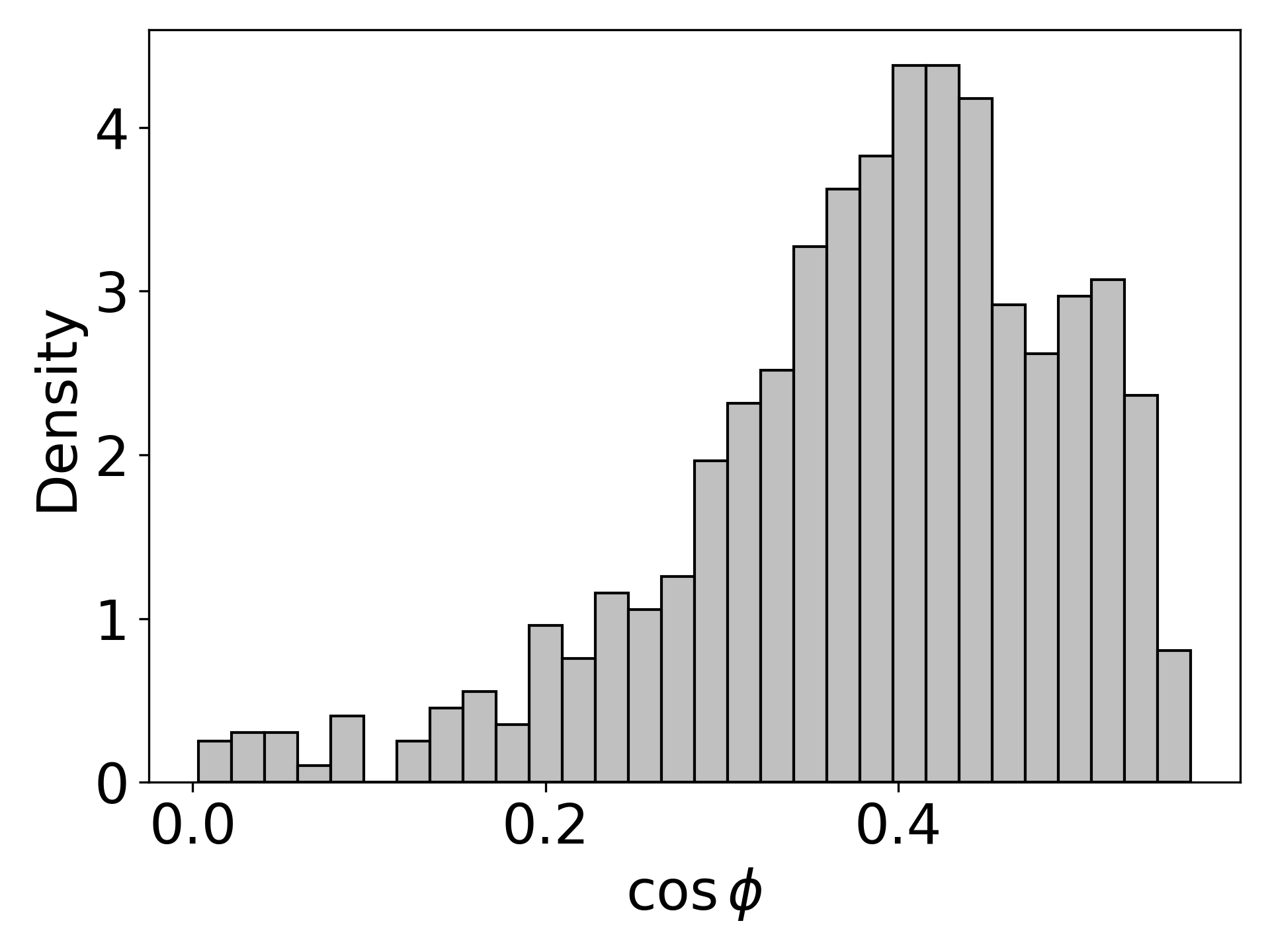}
\caption{}
\label{fig:hist_cosPhi}
\end{subfigure}
\caption{Histograms of (a) the~first Euler angle $\varphi_1$ and (b) $\cos \phi$ within the~transformed fundamental zone \eqref{eq:transformed_fz} for NiTi dataset.}
\label{fig:hist_eu}
\end{figure}

Figure~\ref{fig:cross_sections_NiTi} indicates that $\varphi_1$ and $(\phi,\varphi_2)$ of the~NiTi dataset are independent, and also that
\begin{equation}
\label{eq:uni}
\mbox{$\varphi_2$ conditioned on $\phi$ is uniformly distributed on $[\arcsin \cot\phi,\arccos \cot\phi]$,}
\end{equation}
where the~endpoints of this interval are obtained by considering \eqref{eq:transformed_fz} and solving the~equation $\phi=\phi_0(\varphi_2)$ with respect to $\varphi_2$. Moreover, Figure~\ref{fig:hist_eu} shows histograms of $\varphi_1$ and $\cos\phi$ based on
the~NiTi dataset. Figure~\ref{fig:hist_phi1} suggests that $\varphi_1$ follows a~multimodal distribution which we do not recognize as a~known distribution. Therefore, we use periodic cubic B-splines (with period $2\pi$) \cite{Boor1972} to estimate the~pdf of $\varphi_1$; this estimate is denoted $f_1(\varphi_1)$. 
Furthermore, Figure~\ref{fig:hist_cosPhi} indicates that $\sqrt 3\cos\phi$ follows a~beta distribution with shape parameters $\alpha>0$ and $\beta>0$. Replacing $(\alpha,\beta)$ by its maximum likelihood estimate $(11.79,2.33)$ and using \eqref{eq:uni}, we obtain an~estimated pdf of $(\phi,\varphi_2)$ which is denoted $f_2(\phi,\varphi_2)$. Thus,
\begin{equation}
\label{eq:fs}
f_{\text{s}}(g)= f_1(\varphi_1)f_2(\phi,\varphi_2),\quad (\varphi_1, \phi, \varphi_2)\in F
\end{equation}
is the~estimated pdf of $(\varphi_1,\phi,\varphi_2)$ (here the~subscript `s' refers to that a~single orientation is considered). 

In summary, our (estimated) semiparametric model is given by that $\g_n$ and $\C_n$ are independent, where $\g_n$ has pdf
\begin{equation}
\label{eq:p_noint}
f_{\text{noint}}(\g_n) =  \prod_{i=1}^n  f_{\text{s}}(g_i)
\end{equation}
(`noint' in the~subscript refers to `no interaction').

\subsection{Models with interaction terms}\label{sec:int}

Since the~presence of a~dependence between $\g_n$ and $\C_n$ was revealed in \cite{Pawlas2020}, an~extension of the~model  \eqref{eq:p_noint} is needed. Below we therefore consider three closely related models which incorporate pairwise interaction between orientations of neighbouring grains and which only differ in the~choice of certain weights.

The general form of the~pairwise interaction pdf we consider is
\begin{equation}
\label{eq:general}
f_{\text{int,w}}(\g_n\mid\C_n)\propto \left(\prod_{i=1}^n  f_{\text{s}}(g_i)\right) \exp \bigg\{  \theta \sum_{i \sim j} w_{ij} \inn(g_i, g_j) \bigg\},
\end{equation}
where $i \sim j$ means that $C_i$ and $C_j$ are neighbours (that is, they share a~face in the~Laguerre tessellation), the~sum $\sum_{i \sim j}\cdots$ is over all such neighbouring pairs, $w$ is the~collection of all weights $w_{ij}$ with $i \sim j$ and $\theta$ is a~real unknown parameter. The~right-hand side in \eqref{eq:general} is an~unnormalized density where the~normalizing constant depends on both $\C_n$ and $\theta$. Moreover, the~weights are 
assumed to be of three types $w^{(k)}$, $k=0,1,2$:
\[
w_{ij}^{(0)}=1,\quad w_{ij}^{(1)}=\min \left( \frac{|C_i|_3}{|C_j|_3}, \frac{|C_j|_3}{|C_i|_3} \right),\quad w_{ij}^{(2)}=\frac{|C_i \cap C_j|_2}{\min\left( |C_i|_2, |C_j|_2\right)},
\]
where for a~Borel set $C\subseteq W$, $|C|_2$ and $|C|_3$ denote the~surface area and volume of $C$, respectively. Thus in \eqref{eq:general}, for $w=w^{(k)}$, 
\begin{itemize}
    \item there is no weighting if $k=0$;
    \item the~weight can range from 0 to 1 when $k>0$;
    \item if $k=1$, high weights mean similar volumes;
    \item if $k=2$, the~weight specifies how large the~surface area of the~common face is as compared to the~smaller surface area of the~two grains. 
\end{itemize}
The weight $w^{(2)}$ may be understood by considering first the~numerator $|C_i \cap C_j|_2$ and then the~denominator $\min\left( |C_i|_2, |C_j|_2\right)$ which is added in order to obtain a~dimensionless quantity. Clearly, $w^{(0)}$ and $w^{(1)}$ are also dimensionless. 
 
Other kinds of interaction models could of course be considered. 
Furthermore, the~inner product in \eqref{eq:general} can be replaced with the~disorientation angle. However, since the~disorientation angle and the~inner product for the~orientations of two neighbouring grains are strongly negatively correlated (cf. Figure~\ref{fig:dis_inn_NiTi}), the~behaviour of such a~model would not be much different. Moreover, simulation under the~model with disorientation angle is much more
time-consuming because we have to consider all 24 elements of $\mathcal{O}$ when calculating the~disorientation angle.

\section{Methods}
\label{sec:methods}

This section describes the~methods used for the~analysis of the~NiTi dataset from Section~\ref{sec:data}.

Our first model has already been fitted in Section~\ref{sec:simple}, but for the~three models in Section~\ref{sec:int} it remains to estimate the~parameter $\theta$. Recall that the~general form of the~pdf for the~three models is given by \eqref{eq:general}. This pdf depends on an~intractable normalizing constant $c(\theta,\C_n)$ given by an~integral with respect to $n$-fold product measure of $\mu_F$, cf.\ Section~\ref{sec:quotient}. 
Therefore we consider a~pseudolikelihood function \cite{Besag1975} for $\theta$ which does not depend on $c(\theta,\C_n)$ as it is given by the~product of the~$n$ full conditional densities for the~$n$ orientations. For $i=1,\dots,n$, denote $(\g_n)_{-i}$ the~vector obtained from $\g_n$ by omitting the~$i$th orientation $g_i$, and $f_{i\theta}(\cdot\mid (\g_n)_{-i},\C_n)$ the~conditional pdf for the~$i$th orientation given the~rest. 
By \eqref{eq:general},
\[
f_{i\theta}(g_i\mid (\g_n)_{-i},\C_n)\propto f_{\mathrm s}(g_i)\exp\left\{\theta\sum_{j:\,i\sim j}w_{ij}\inn(g_i,g_j)\right\},
\]
where the~normalizing constant is
\[
c_i(\theta,\C_n) = \int f_{\mathrm s}(g_i)\exp\left\{\theta\sum_{j:\,i\sim j}w_{ij}\inn(g_i,g_j)\right\}\,\mathrm d\mu_F(g_i).
\]
Now, the~log-pseudolikelihood function for $\theta$ is
\[
l(\theta)=\sum_{i=1}^n\ln f_{i\theta}(g_i\mid (\g_n)_{-i},\C_n).
\]
The first and second derivatives of $l(\theta)$ are easily derived and it follows that $l(\theta)$  is a~concave function. Hence $l(\theta)$ is easily maximized using the~Newton--Raphson method. Since  $c_i(\theta,\C_n)$ has to be calculated by numerical methods, approximations of the~derivatives $l'(\theta)$ and $l''(\theta)$ are given in \ref{sec:MPL}. 

When inspecting each of the~four fitted models, it is worth stressing that due to the~complexity of the~data in hand we do not make any formal hypothesis tests, such as a~global envelope test \cite{Myllymaki2017}, as we admit that these may very well lead to very small \mbox{$p$-values}. On the~other hand, we concentrate on comparing how well the~empirical distribution of selected orientation characteristics (the orientations themselves as well as those in Section~\ref{sec:chars}) agree with the~corresponding distributions obtained by simulations under a~fitted model. 
This comparison is done partly by plotting the~empirical distribution function for the~data and the~simulations and partly by calculating mean and standard deviations as detailed in Section~\ref{sec:results}.
Moreover, we compare different fitted models by considering the~maximal value of the~log-pseudolikelihood function.

It remains to discuss how to make simulations. For the~first model in Section~\ref{sec:simple}, the~orientations are independent and each orientation follows the~density \linebreak $f_{\mathrm s}(g)=f_1(\varphi_1)f_2(\phi,\varphi_2)$, cf.\ \eqref{eq:fs}. We use rejection sampling when simulating from $f_1$ and a~two-step method for simulation of $(\phi,\phi_2)\sim f_2$: first $\cos \phi$ is sampled from the~fitted beta distribution and then {$\varphi_2$} is sampled from the~conditional uniform distribution, cf.\ \eqref{eq:uni}. 
For the~pairwise interaction models which are given by \eqref{eq:general}, we use a~Metropolis-within-Gibbs algorithm with a~cyclic updating scheme which updates the~orientations from one end to the~other (a so-called sweep, see \cite[Section 4.3.3]{MCMC2011}). Here, when updating the~$i$th orientation, if $g_{i,\mathrm{old}}$ and $(\g_n)_{-i}$ specify the~current state of orientations, we propose a~new value $g_i$ generated from $f_{\mathrm s}$ and accept this with probability $\min\{1,H_i(g_i,g_{i,\mathrm{old}},(\g_n)_{-i})\}$ (otherwise we keep $g_{i,\mathrm{old}}$), where
\[
H_i(g_i,g_{i,\mathrm{old}},(\g_n)_{-i})=
\exp\left\{\theta\sum_{j:\,i\sim j}w_{ij}\left[\inn(g_i,g_j)-\inn(g_{i,\mathrm{old}},g_j)\right]\right\}
\]
is the~Hastings ratio.

\section{Analysis of the NiTi dataset}
\label{sec:results}

\begin{figure*}[t!]
   \centering
\begin{tabular}{cccc}
$f_{\text{noint}}$ &  $f_{\text{int}, w^{(0)}}$ &  $f_{\text{int}, w^{(1)}}$ &  $f_{\text{int}, w^{(2)}}$ \\
    \includegraphics[width=0.22\linewidth]{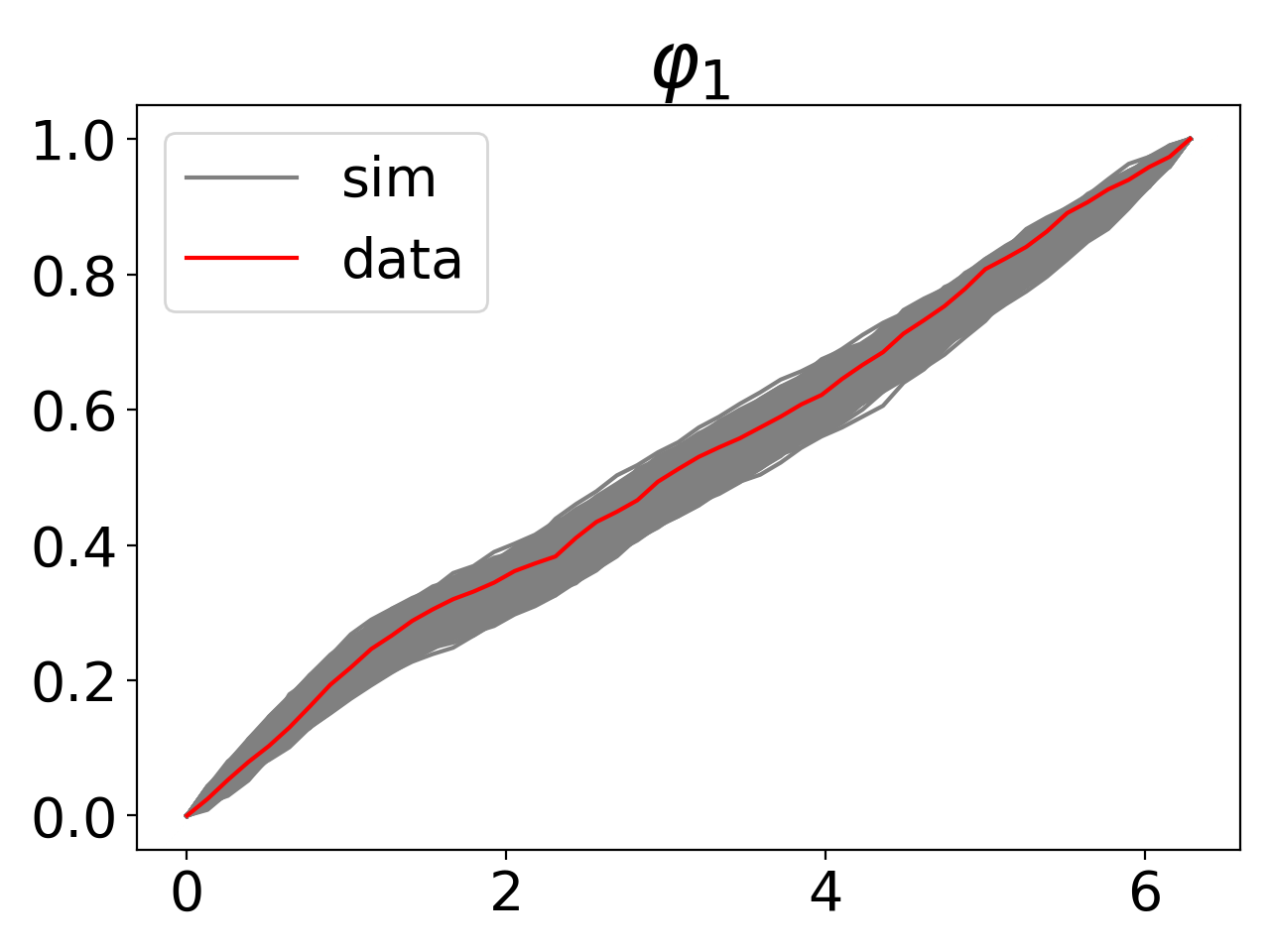} &
    \includegraphics[width=0.22\linewidth]{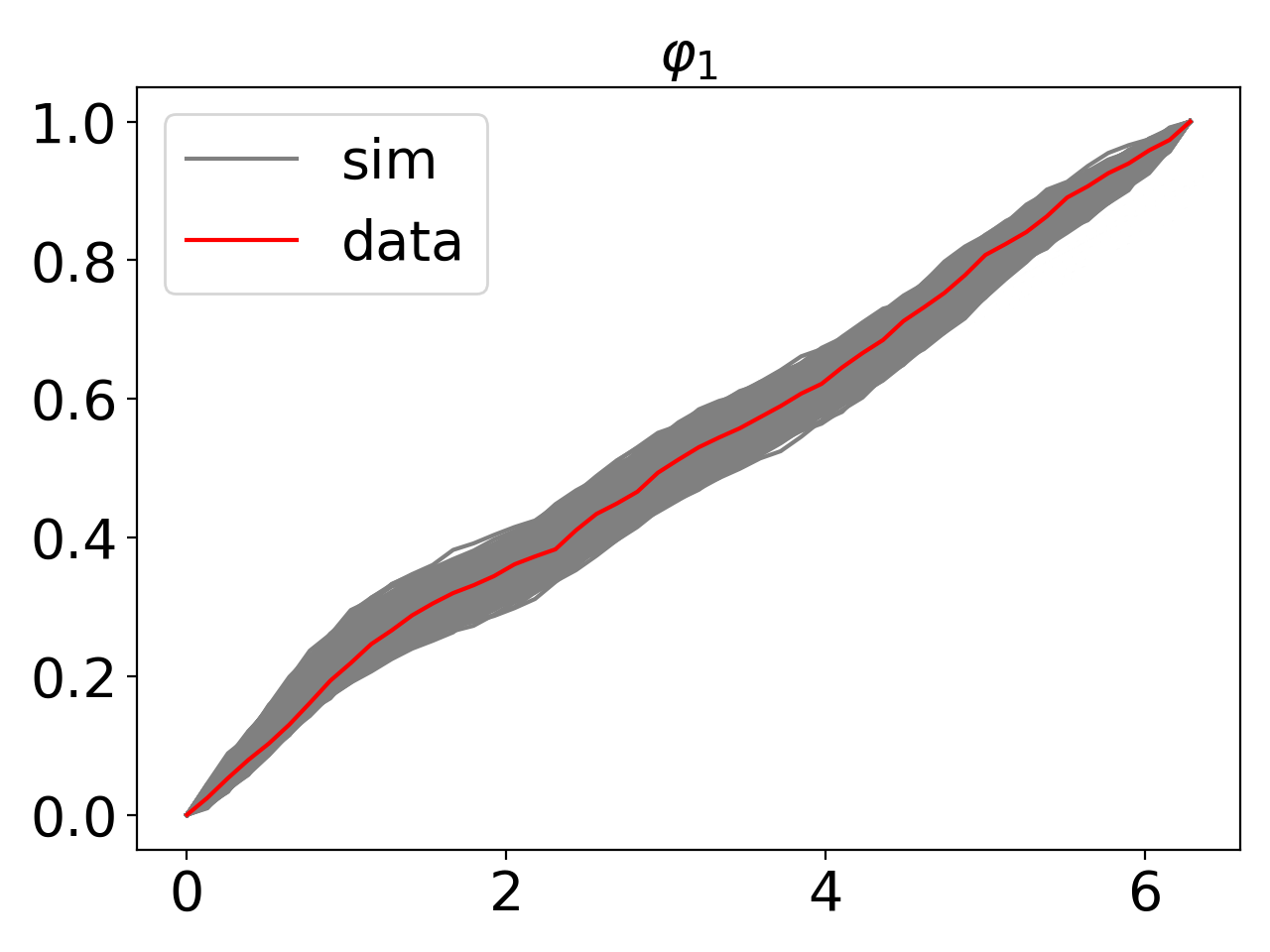} &
    \includegraphics[width=0.22\linewidth]{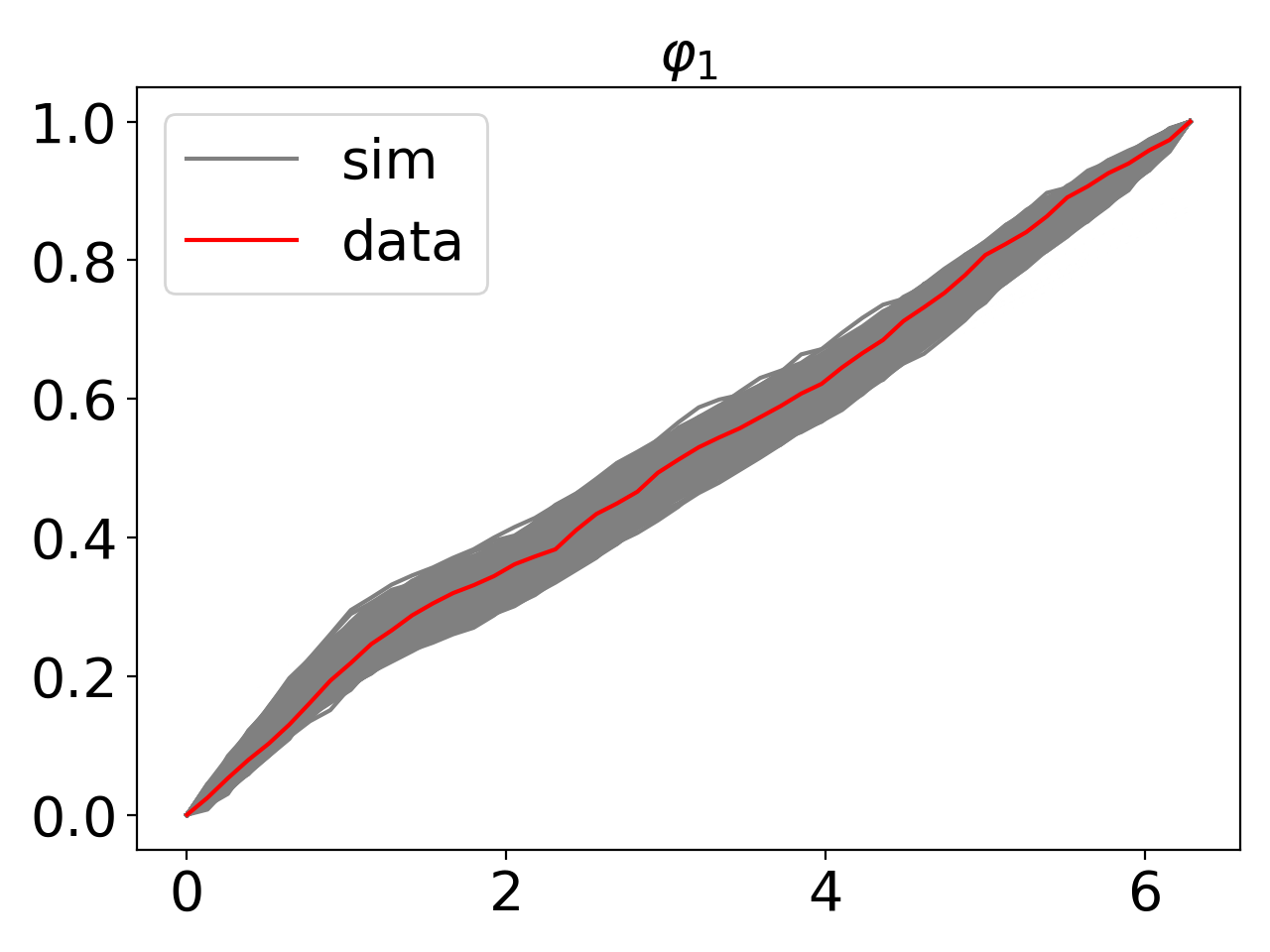} &
    \includegraphics[width=0.22\linewidth]{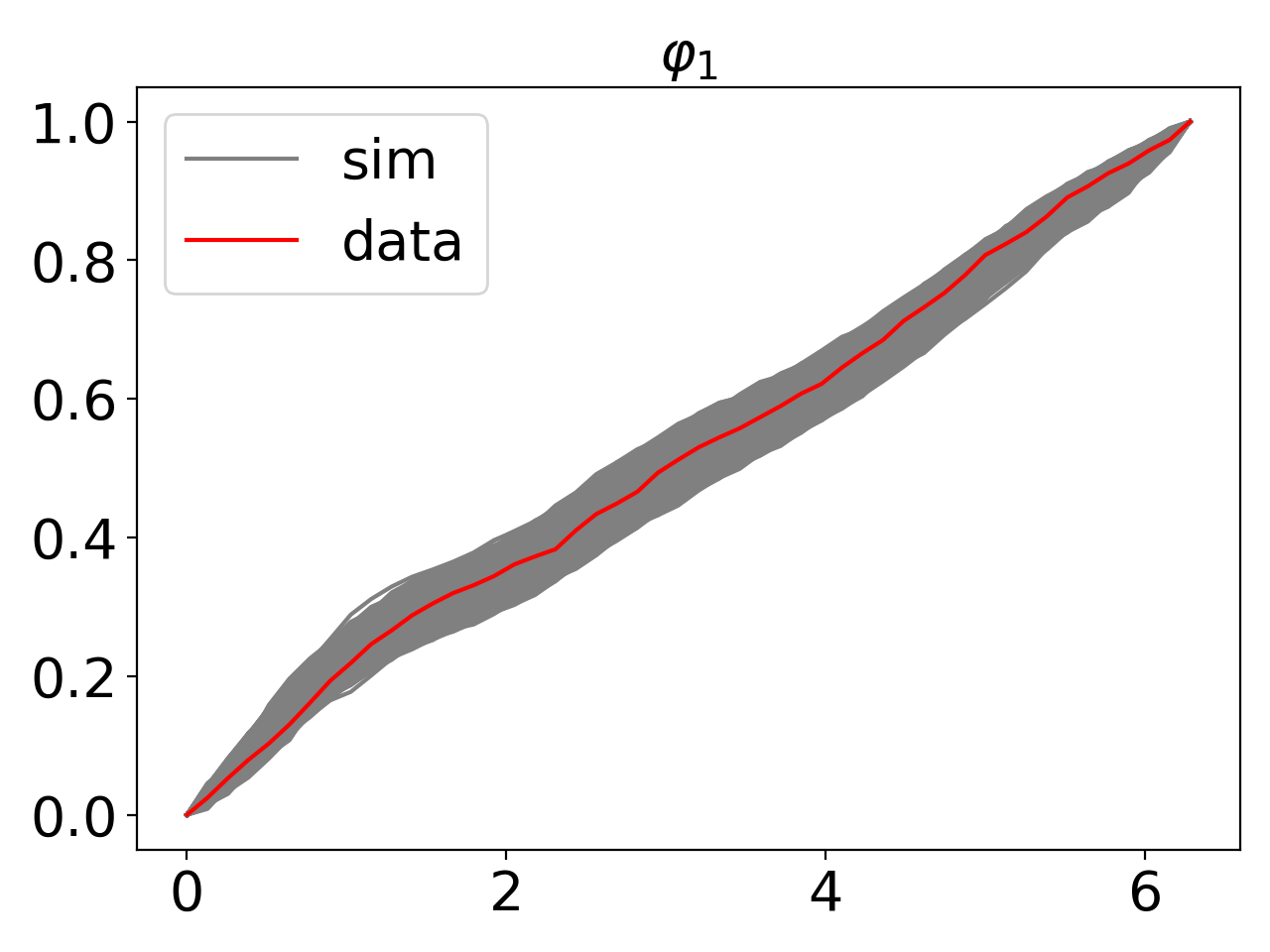} \\
    \includegraphics[width=0.22\linewidth]{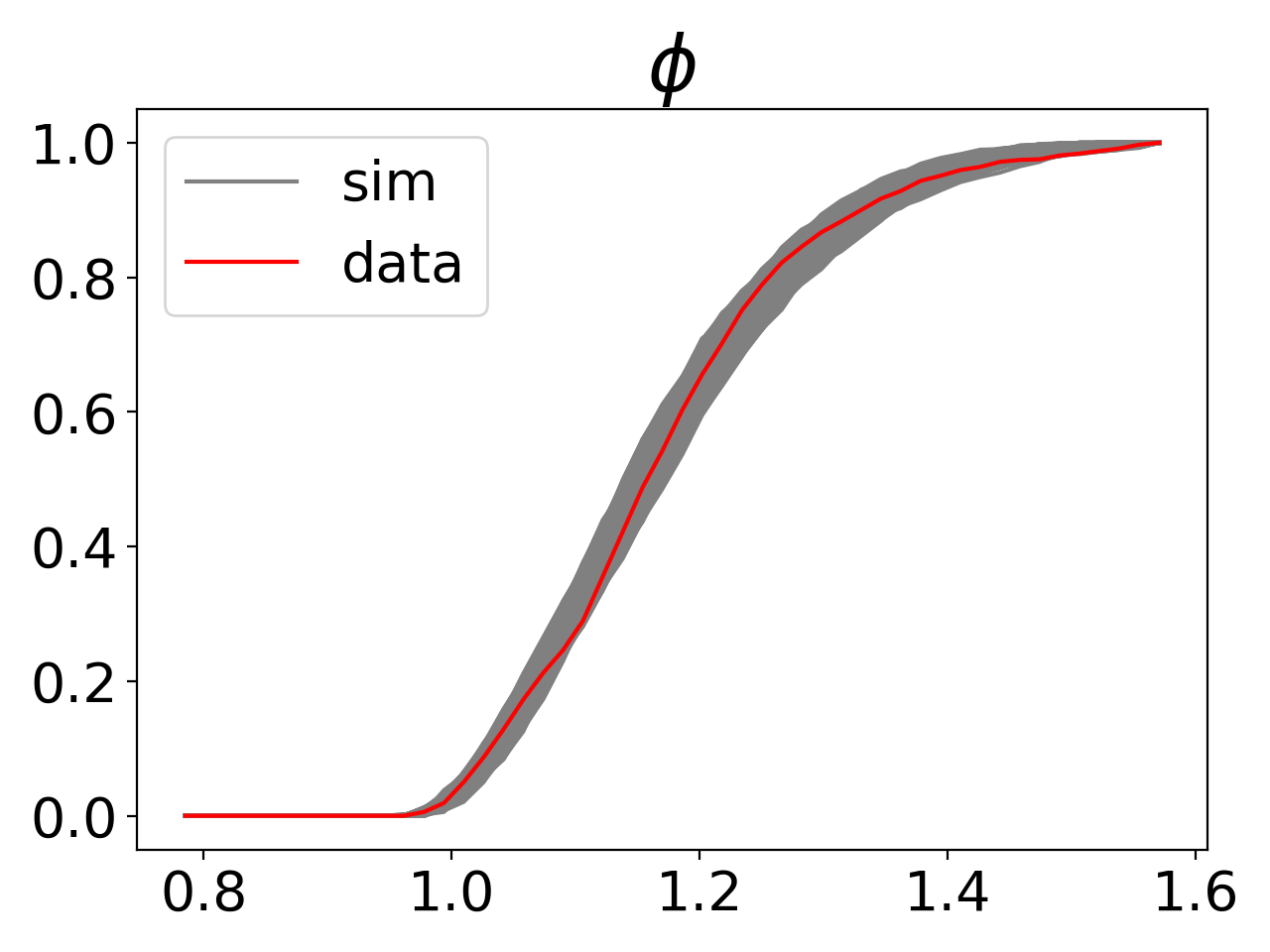} &
    \includegraphics[width=0.22\linewidth]{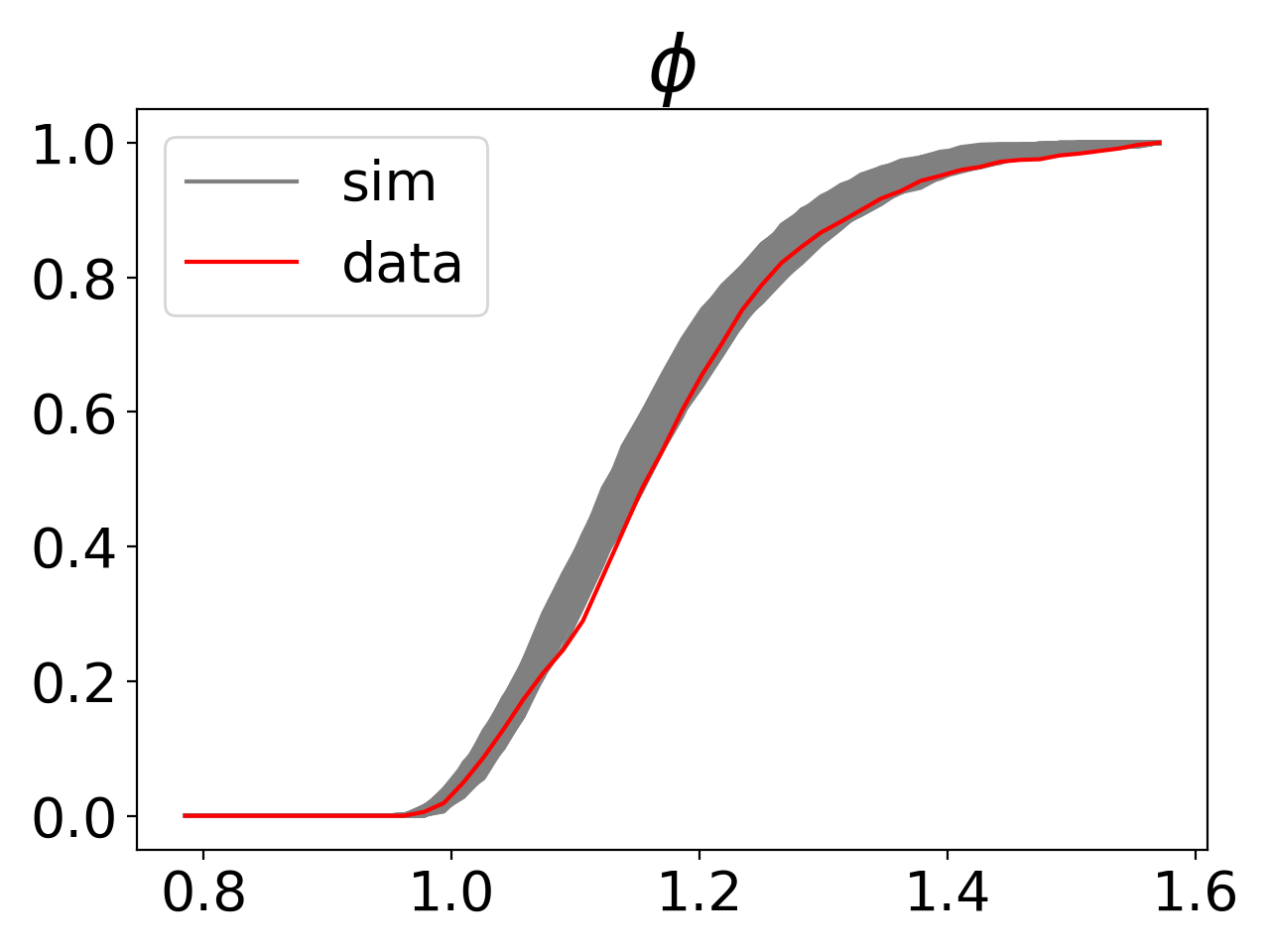} &
    \includegraphics[width=0.22\linewidth]{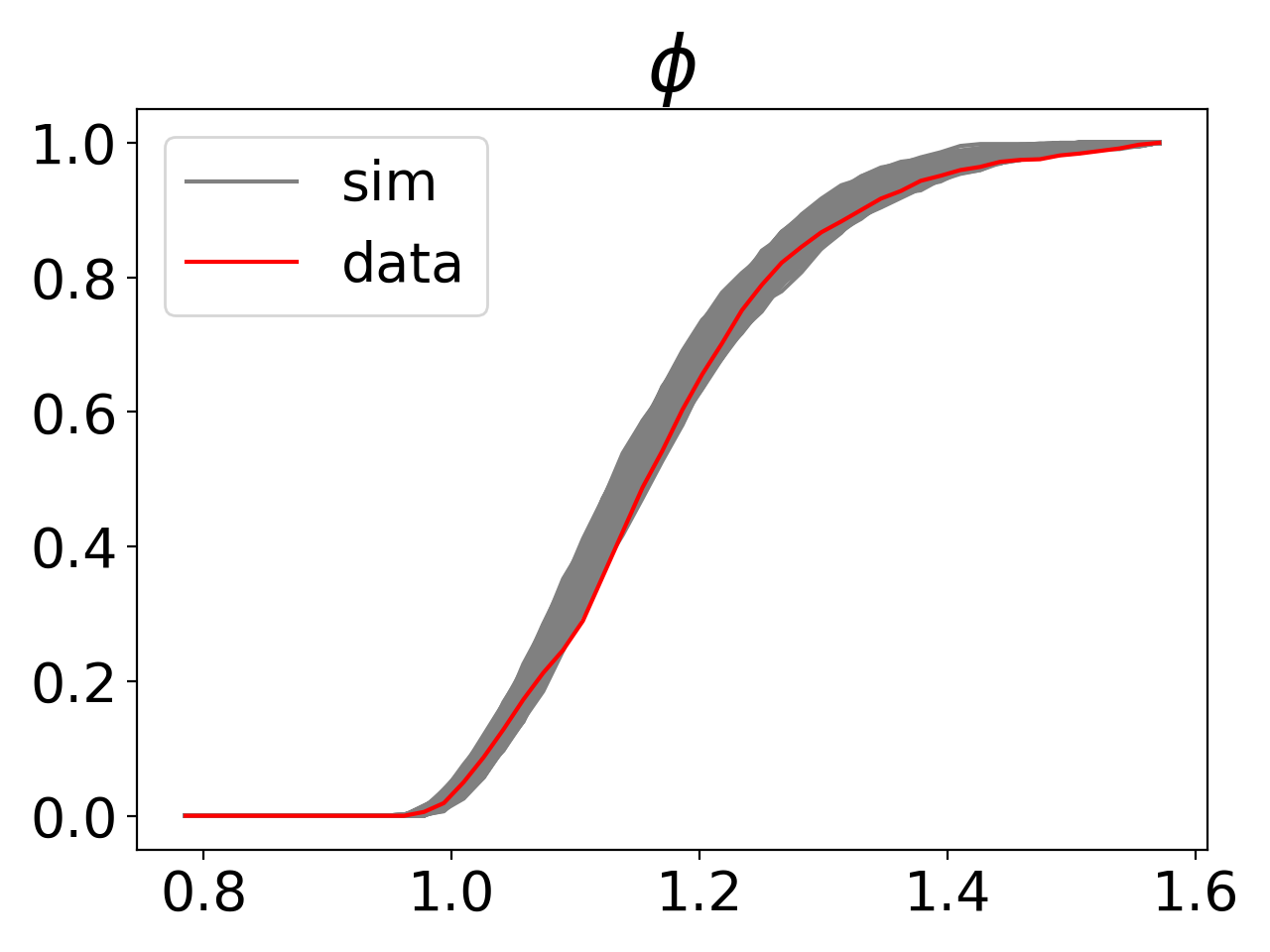} &
    \includegraphics[width=0.22\linewidth]{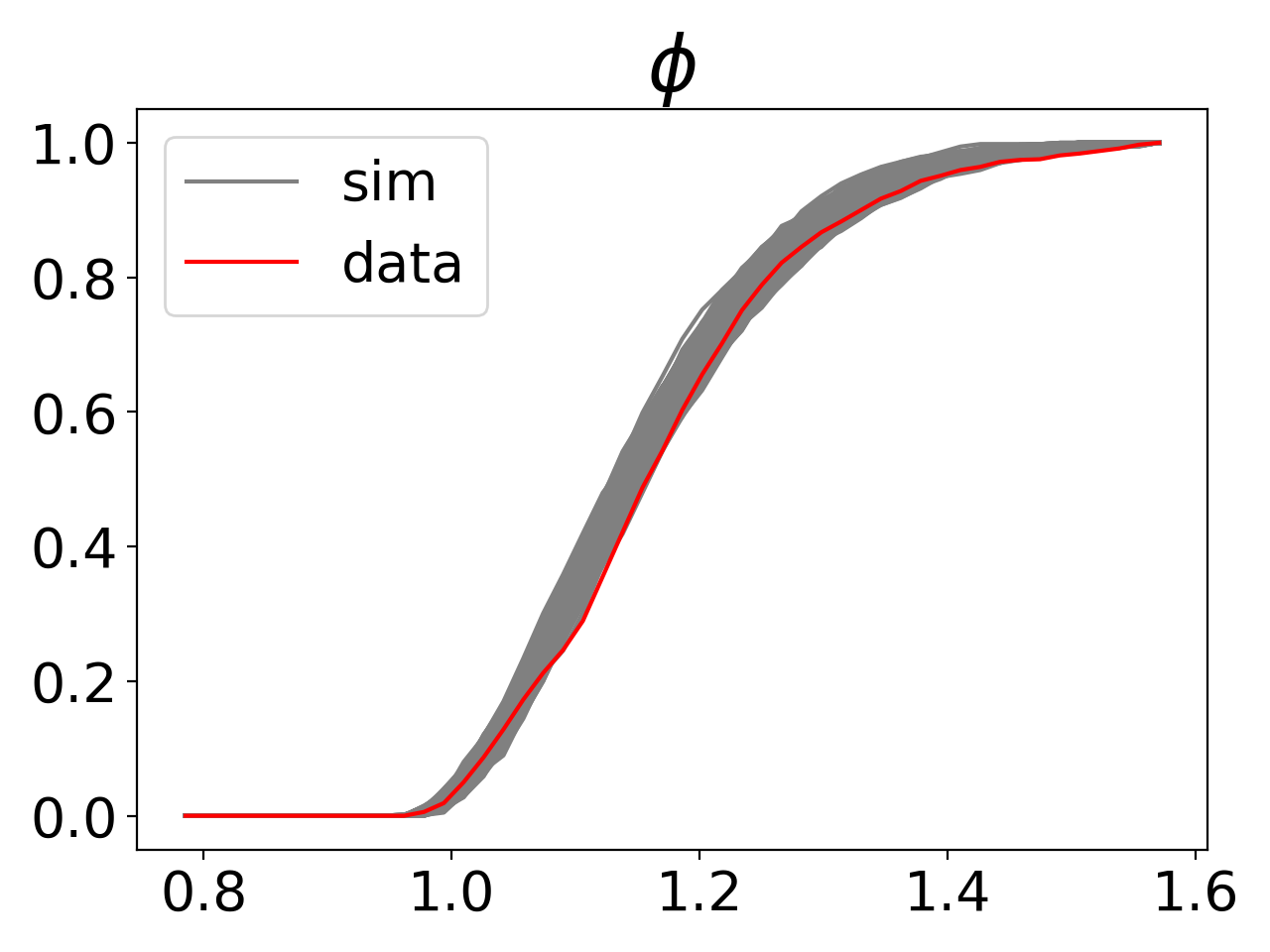} \\
    \includegraphics[width=0.22\linewidth]{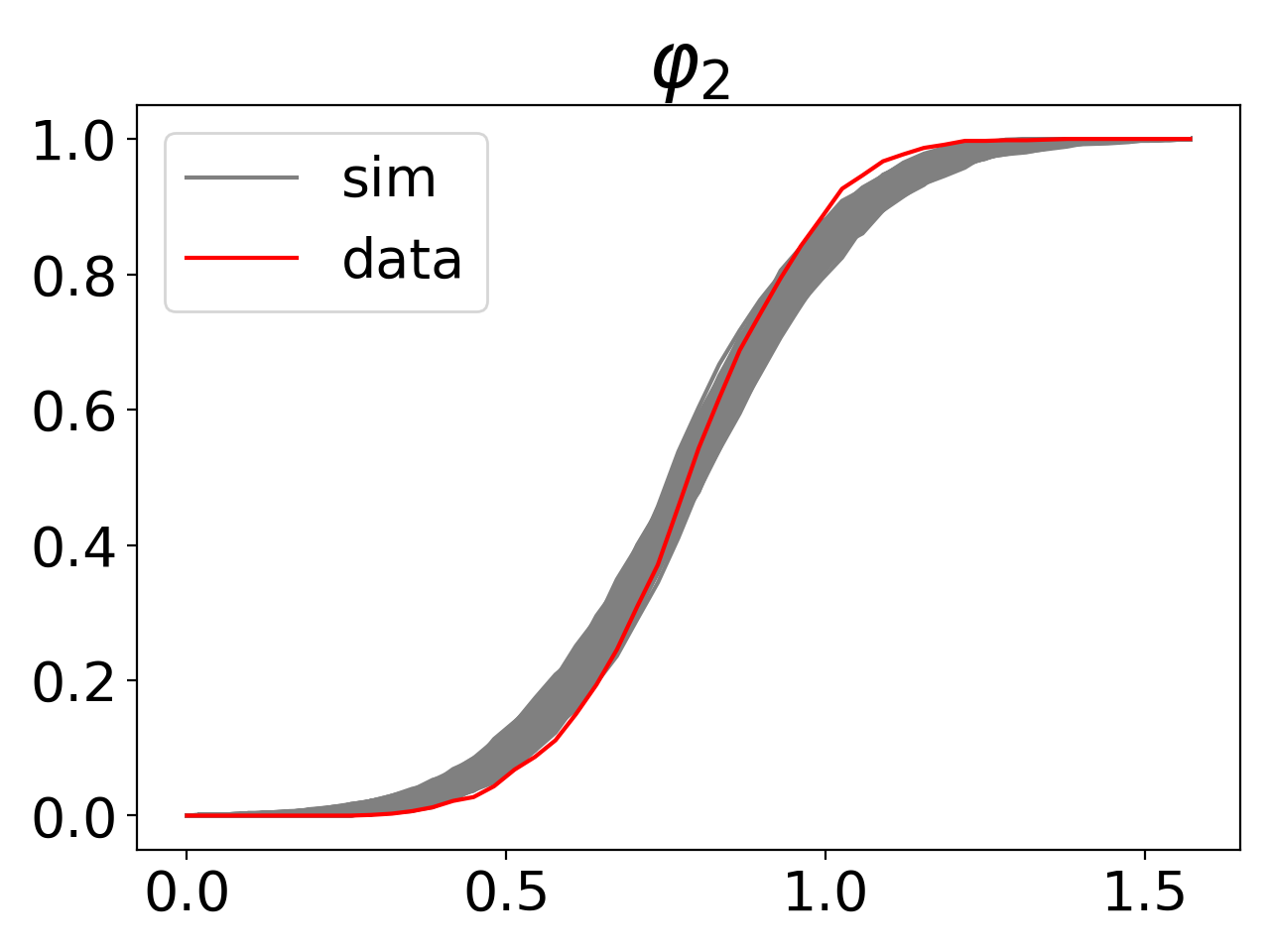} &
    \includegraphics[width=0.22\linewidth]{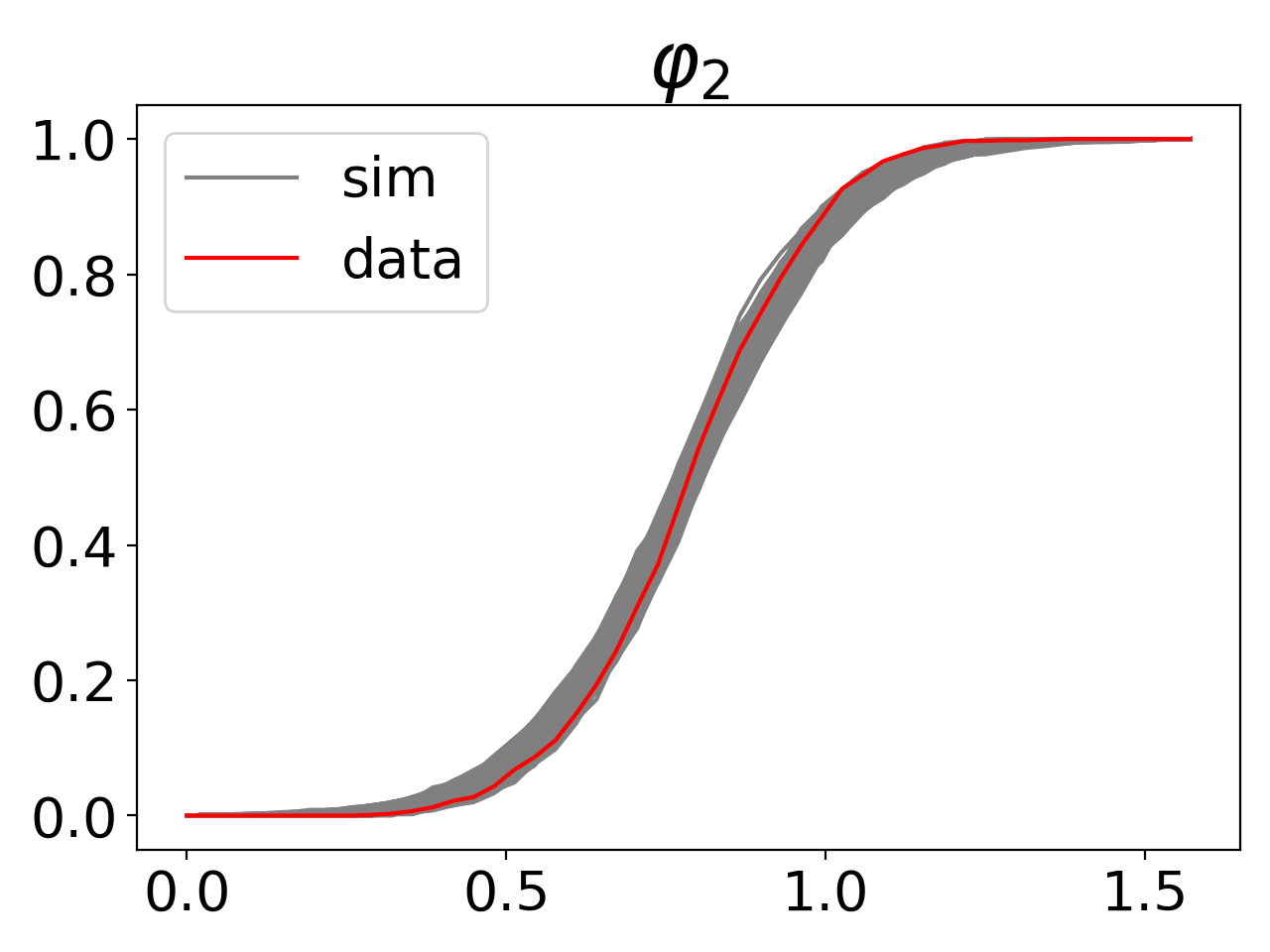} &
    \includegraphics[width=0.22\linewidth]{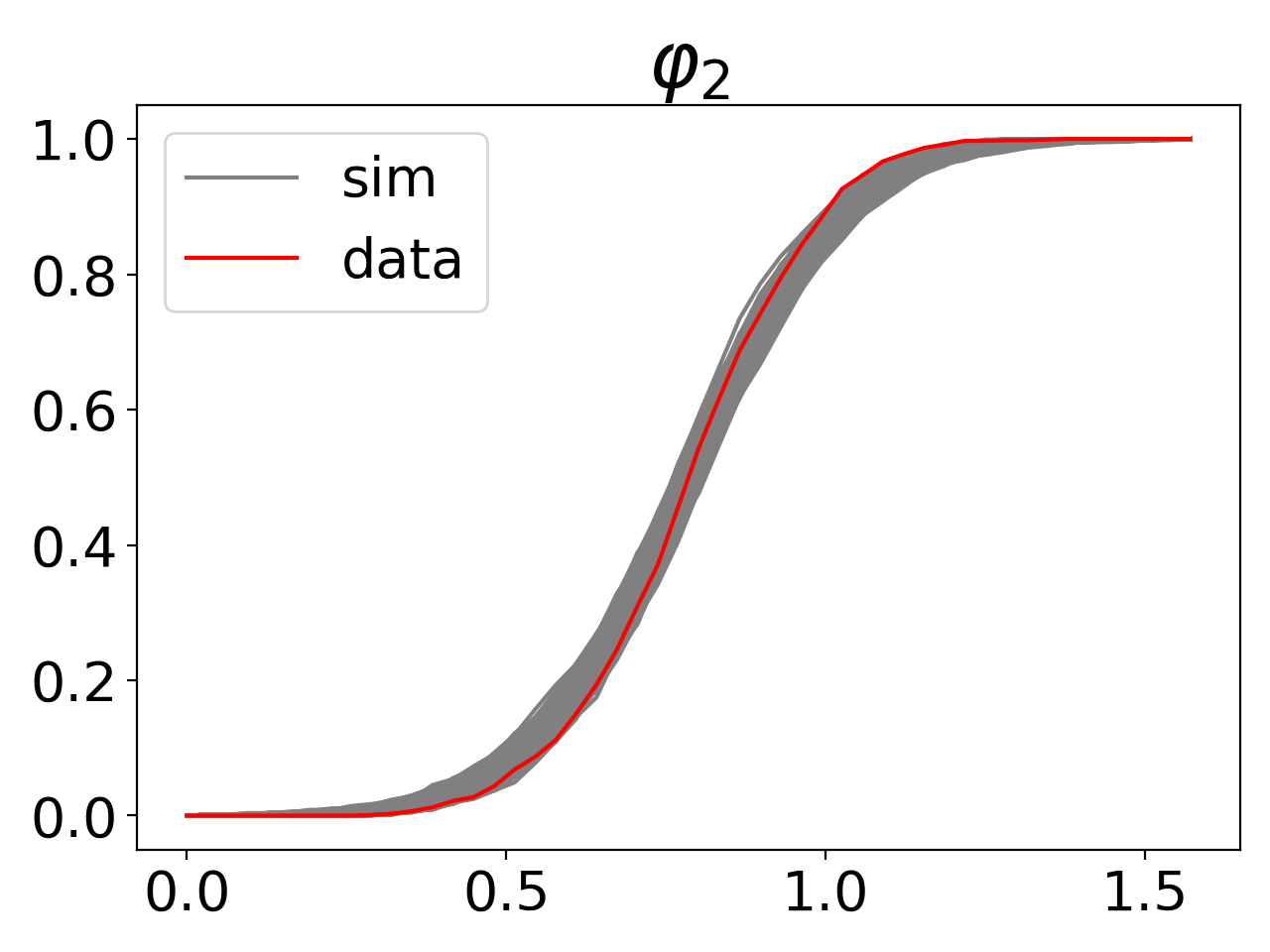} &
    \includegraphics[width=0.22\linewidth]{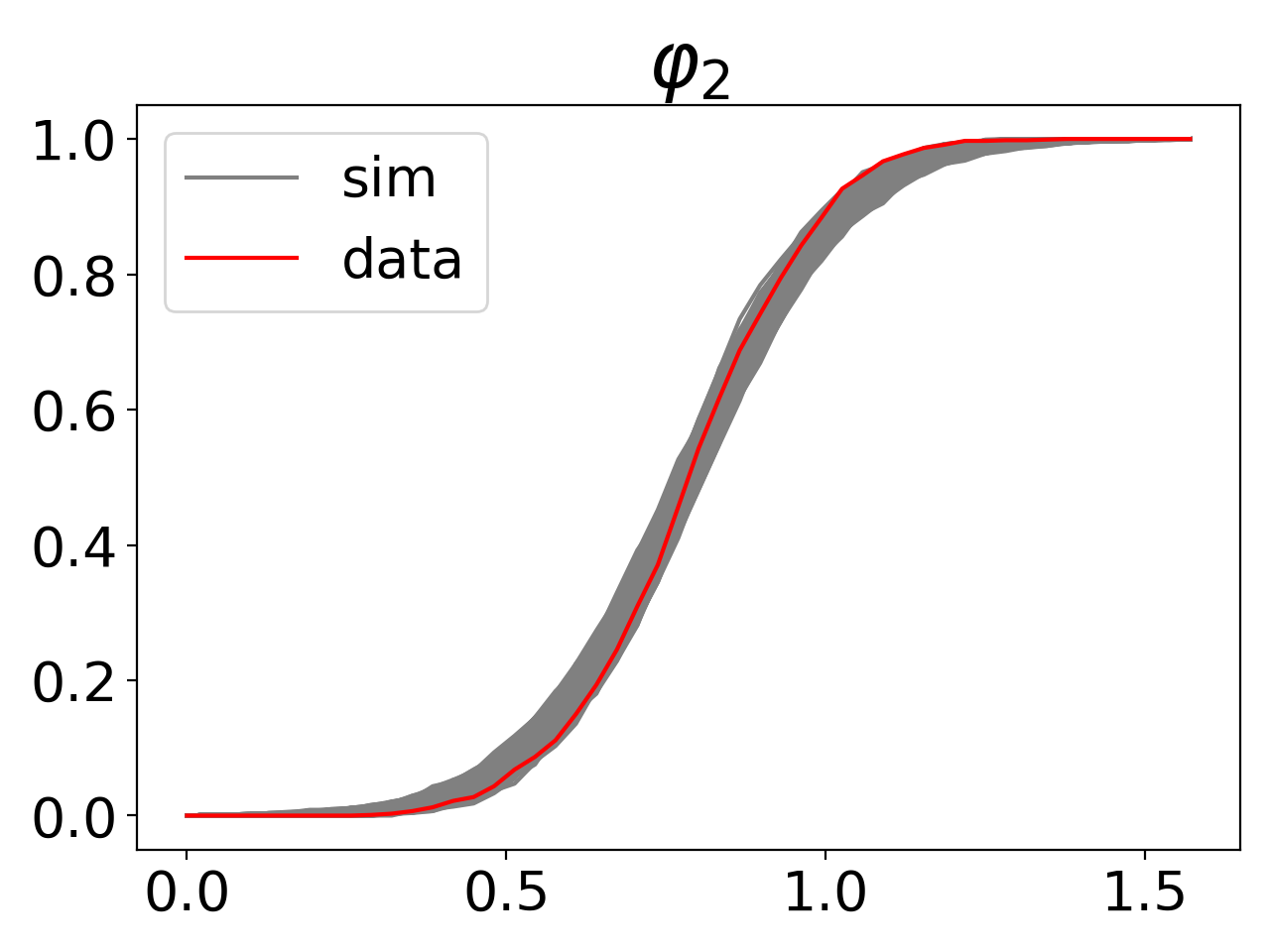} \\
    \includegraphics[width=0.22\linewidth]{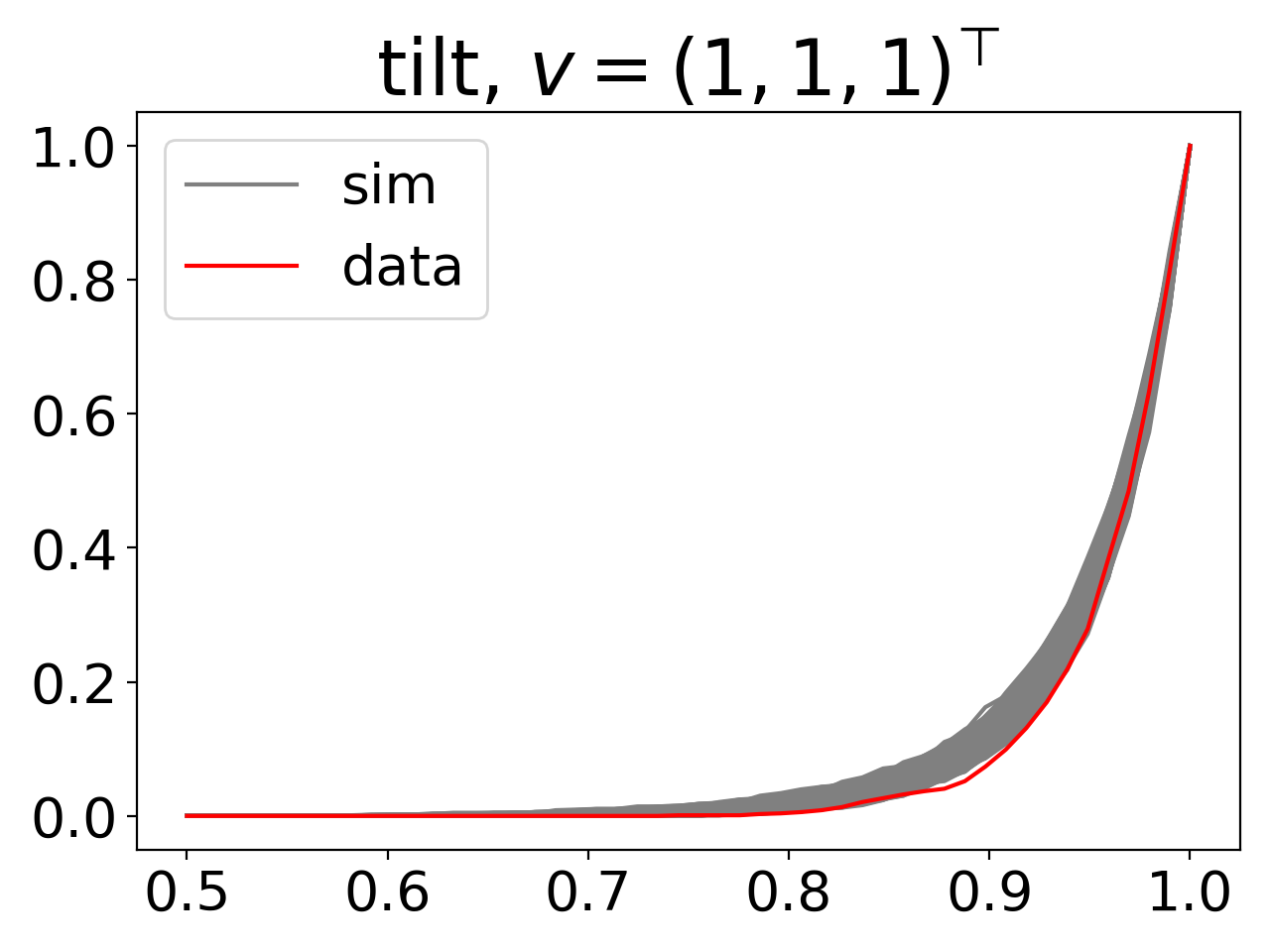} &
    \includegraphics[width=0.22\linewidth]{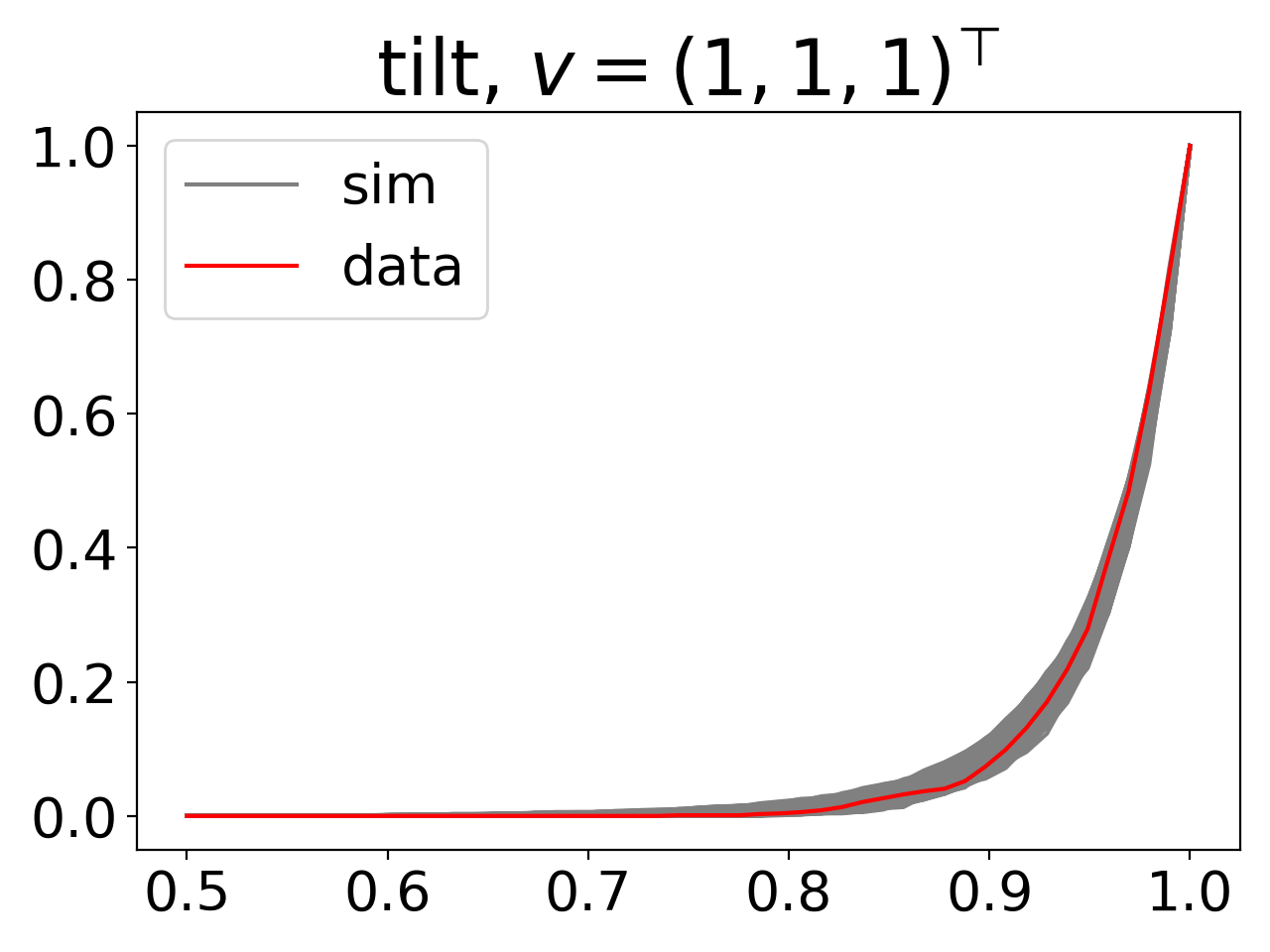} &
    \includegraphics[width=0.22\linewidth]{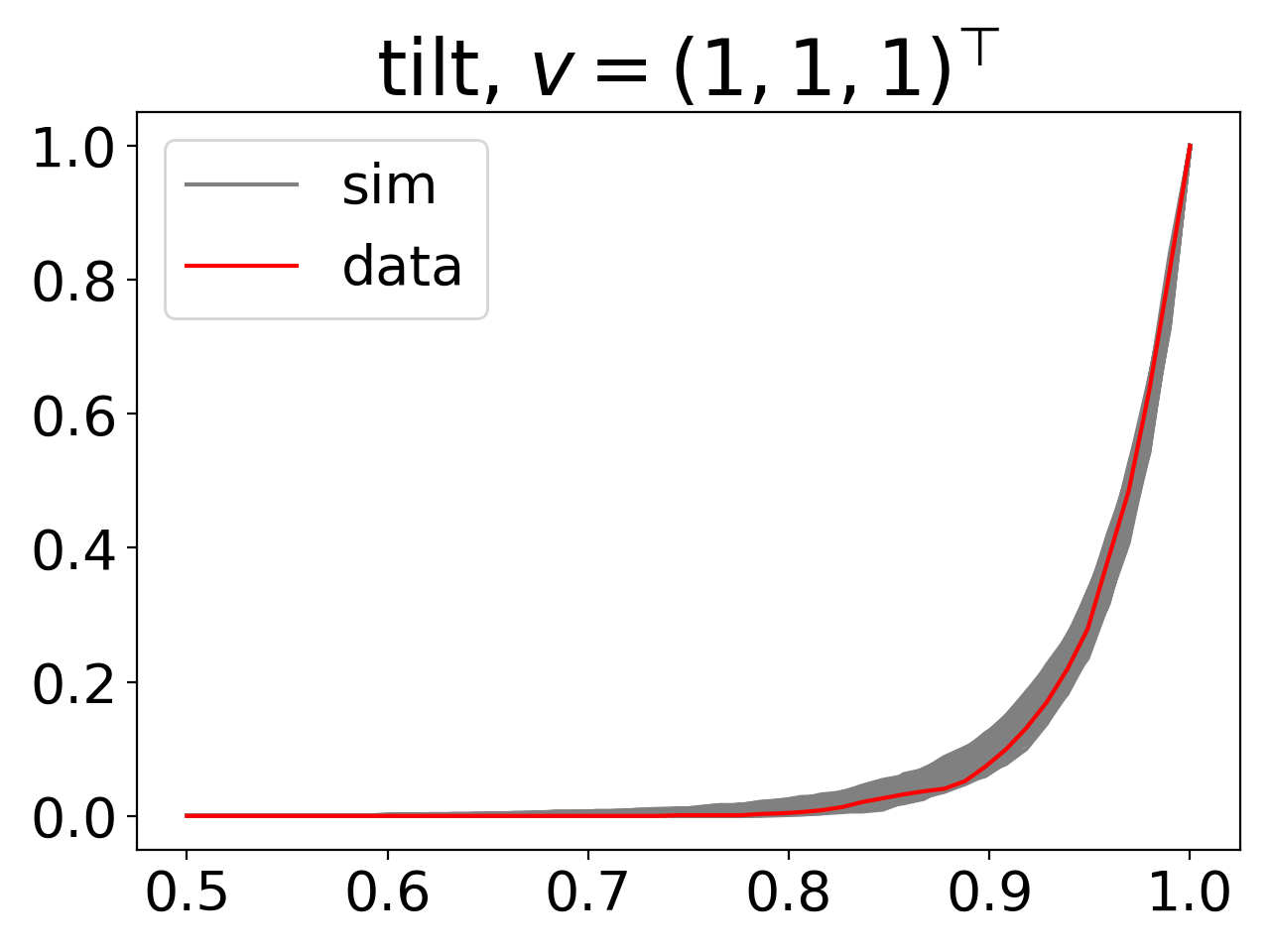} &
    \includegraphics[width=0.22\linewidth]{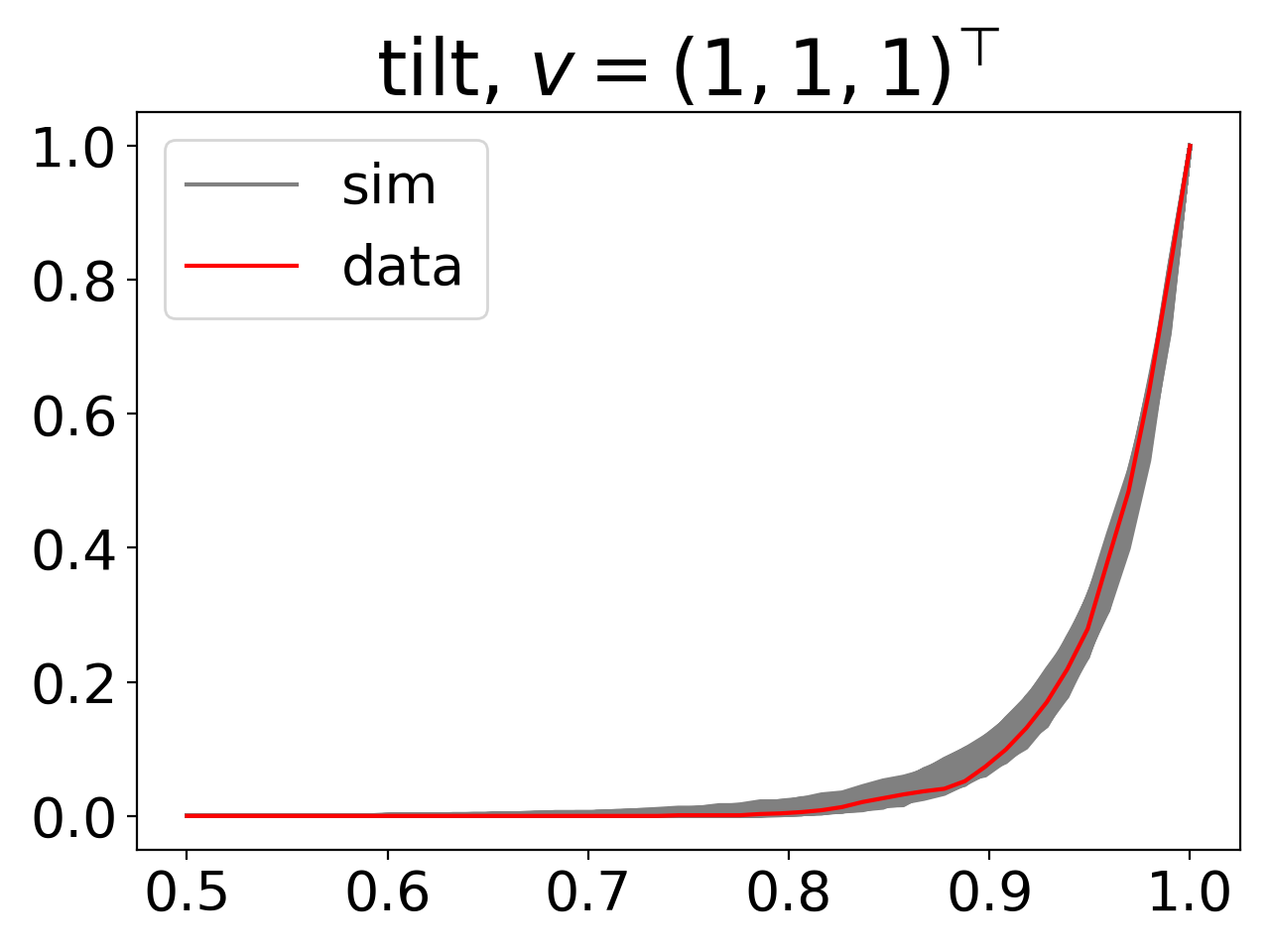} \\
    \includegraphics[width=0.22\linewidth]{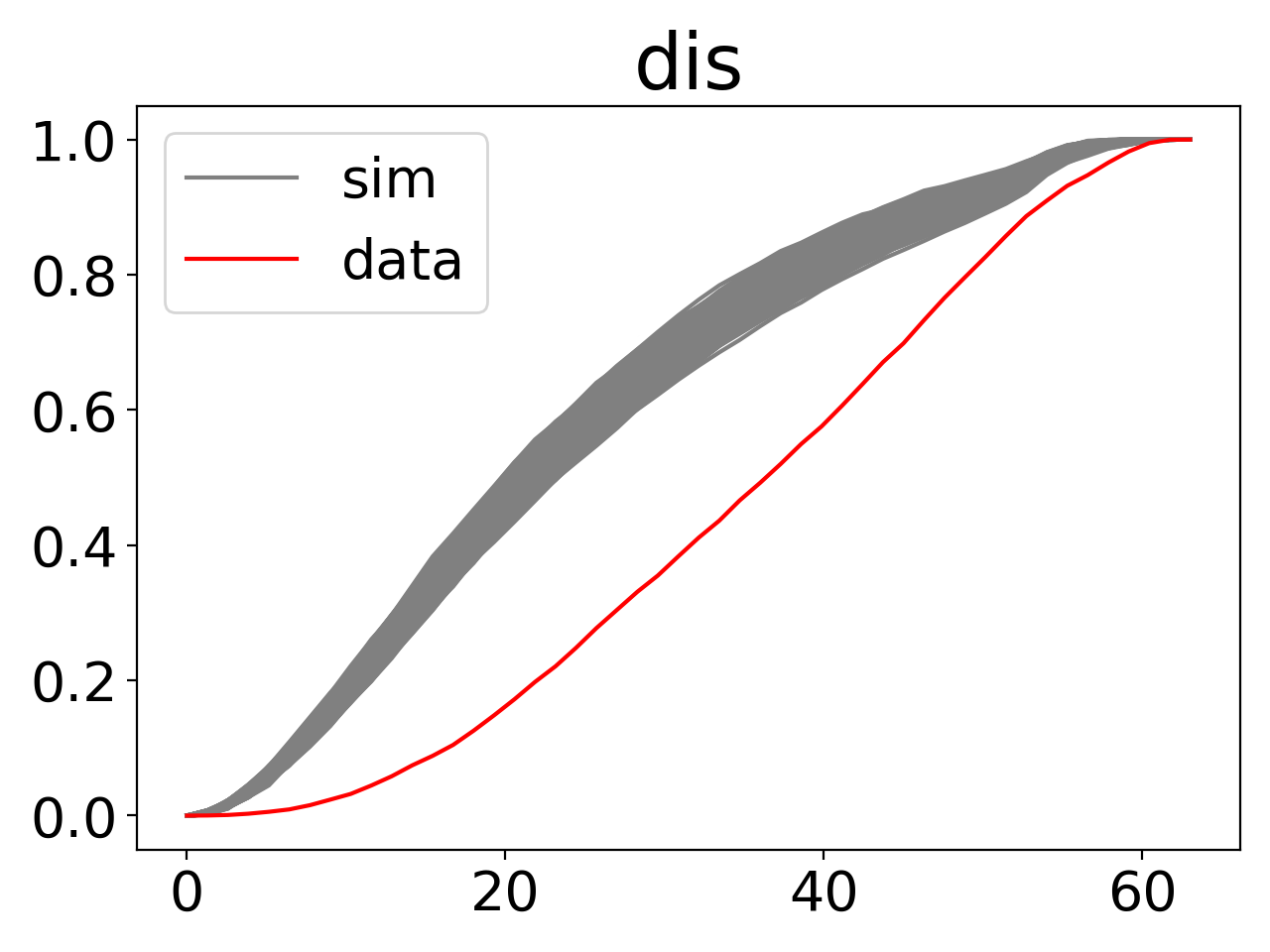} &
    \includegraphics[width=0.22\linewidth]{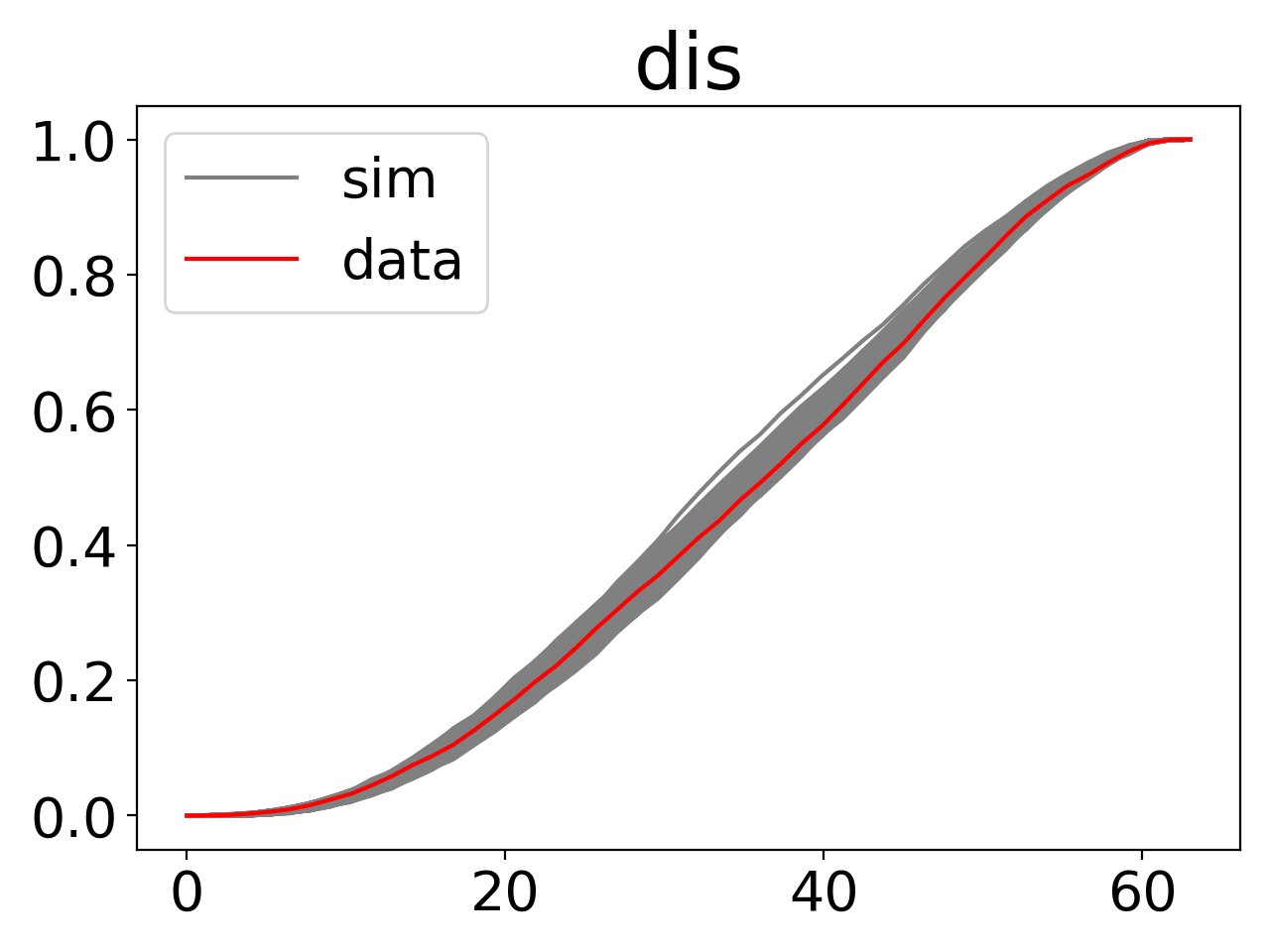} &
    \includegraphics[width=0.22\linewidth]{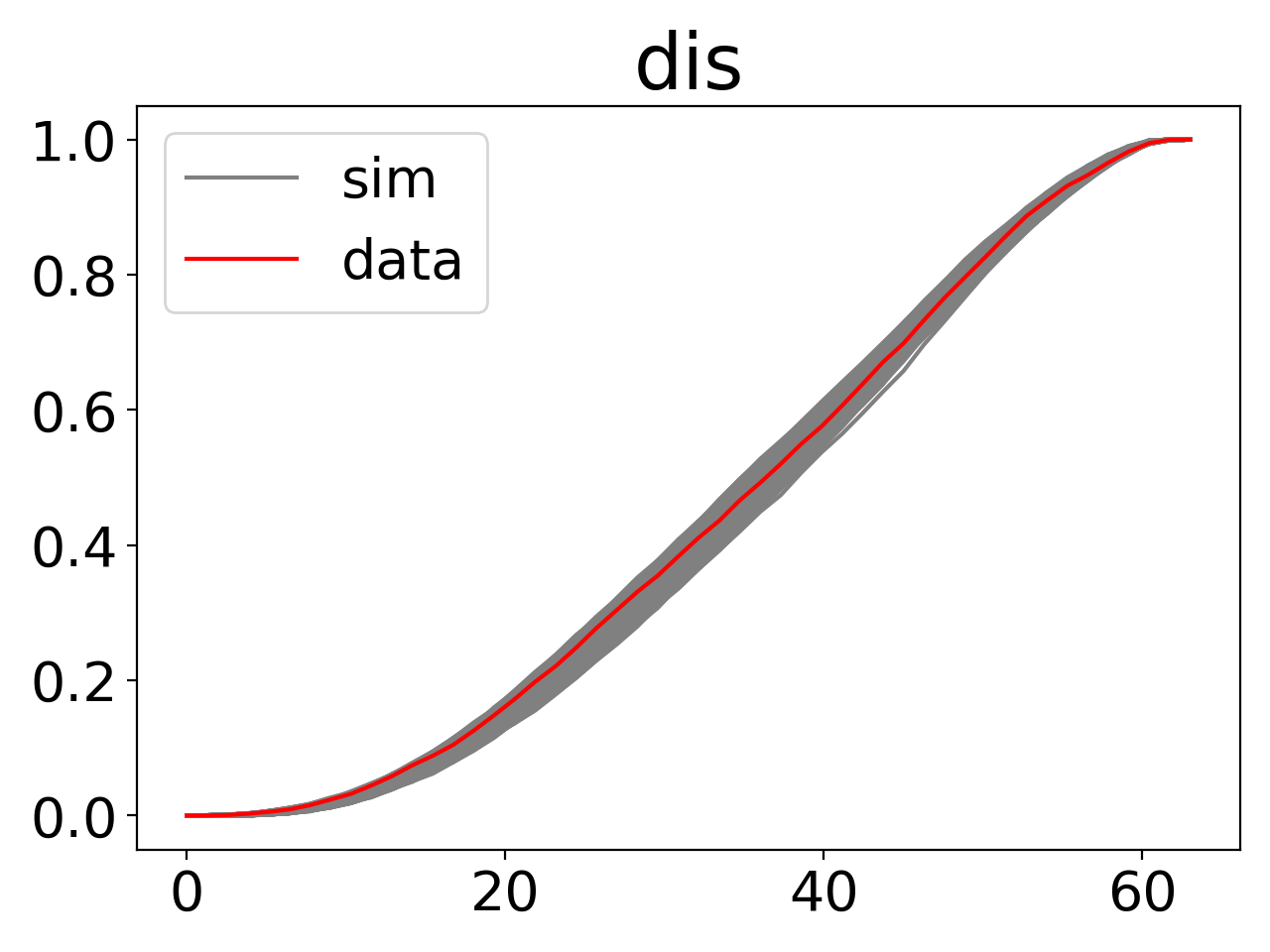} &
    \includegraphics[width=0.22\linewidth]{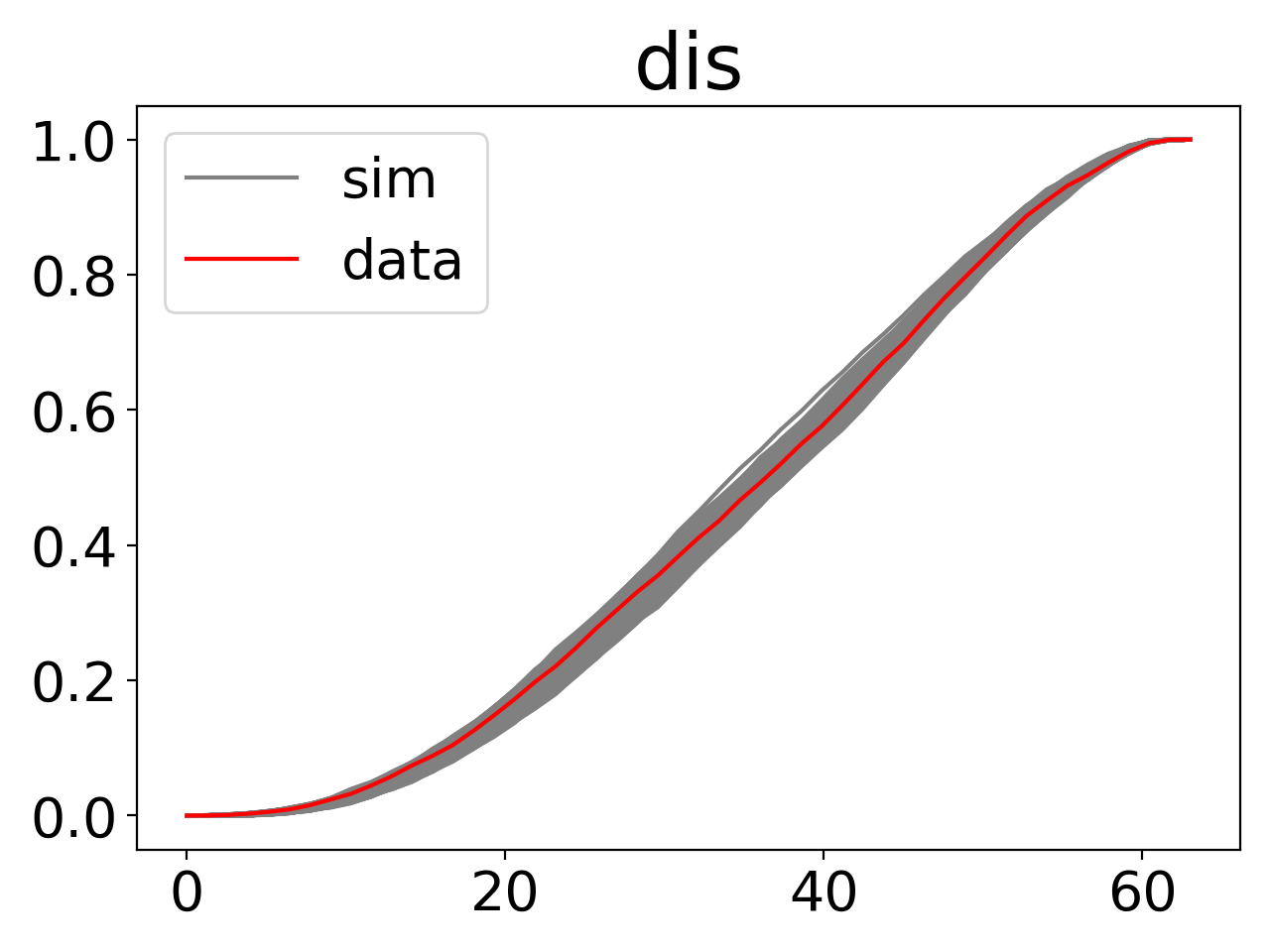} \\
    \includegraphics[width=0.22\linewidth]{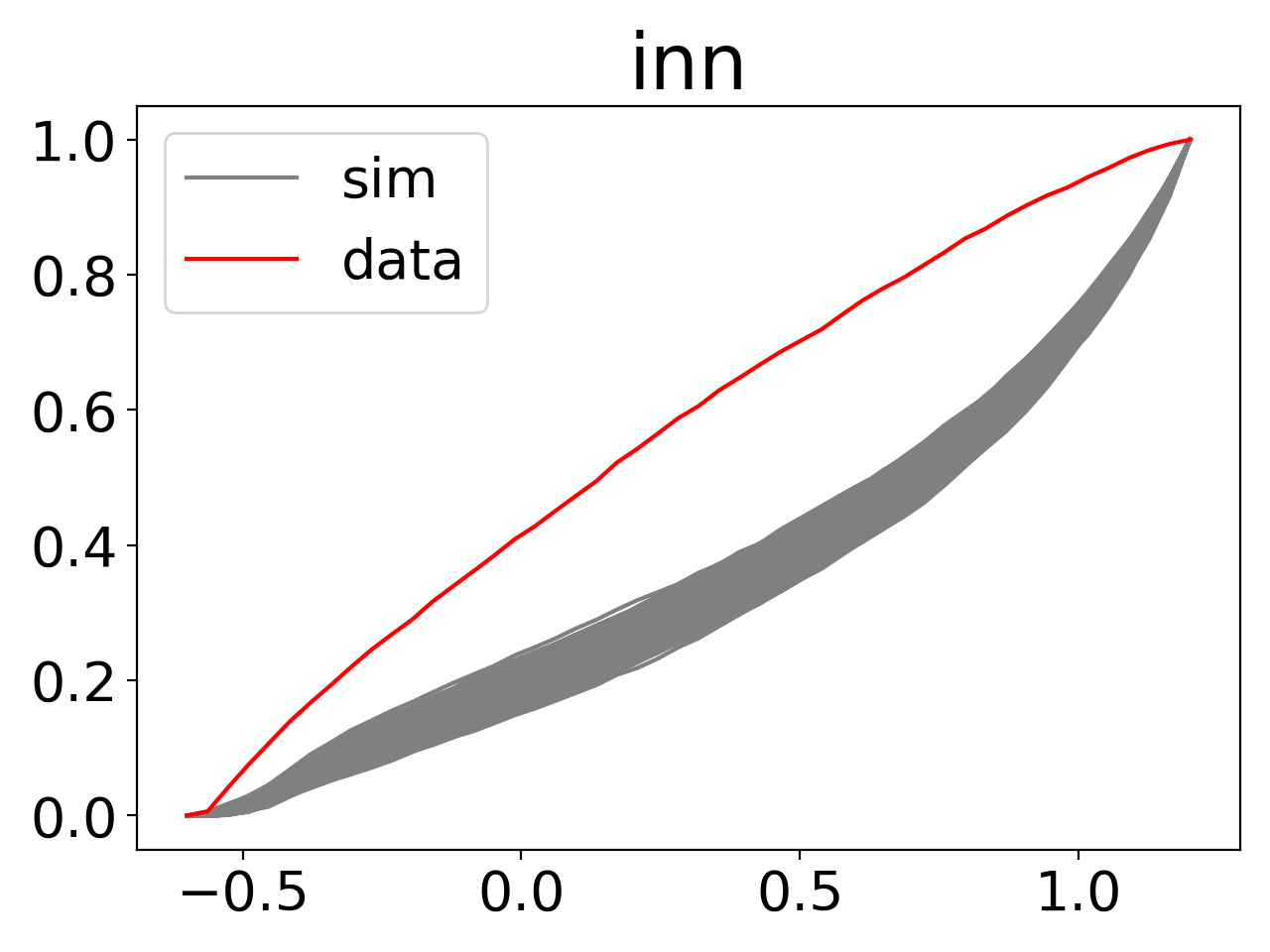} &
    \includegraphics[width=0.22\linewidth]{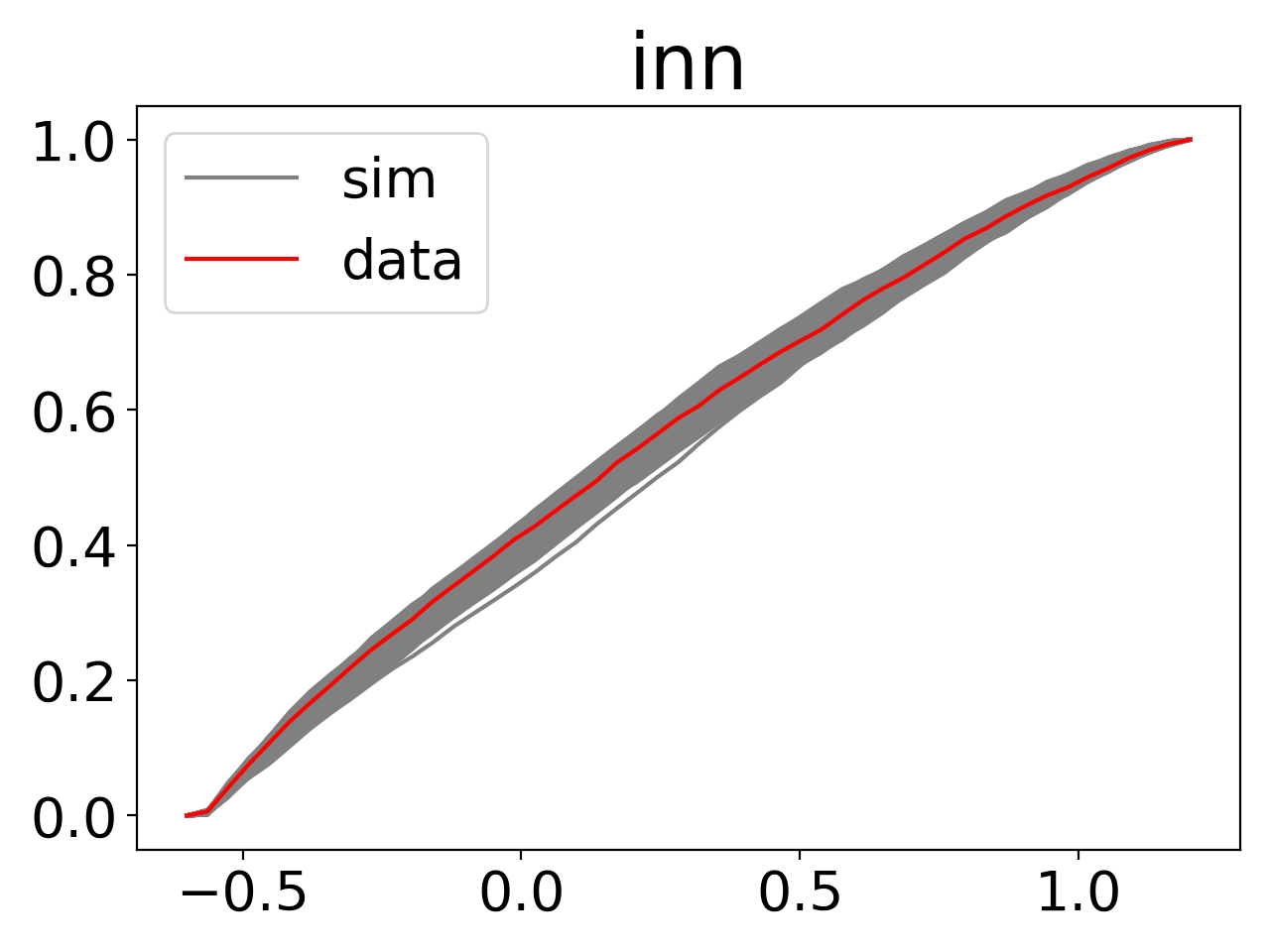} &
    \includegraphics[width=0.22\linewidth]{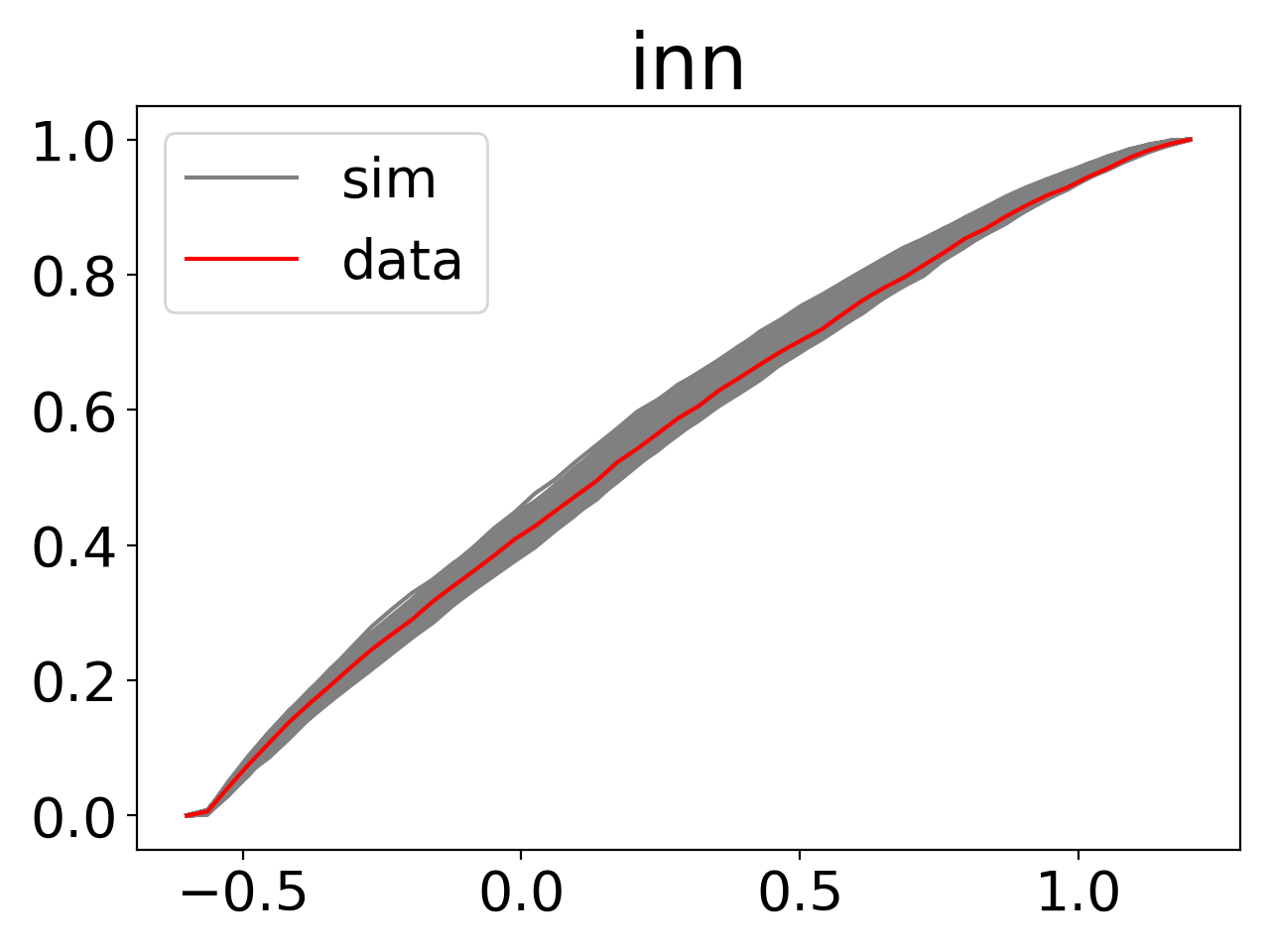} &
    \includegraphics[width=0.22\linewidth]{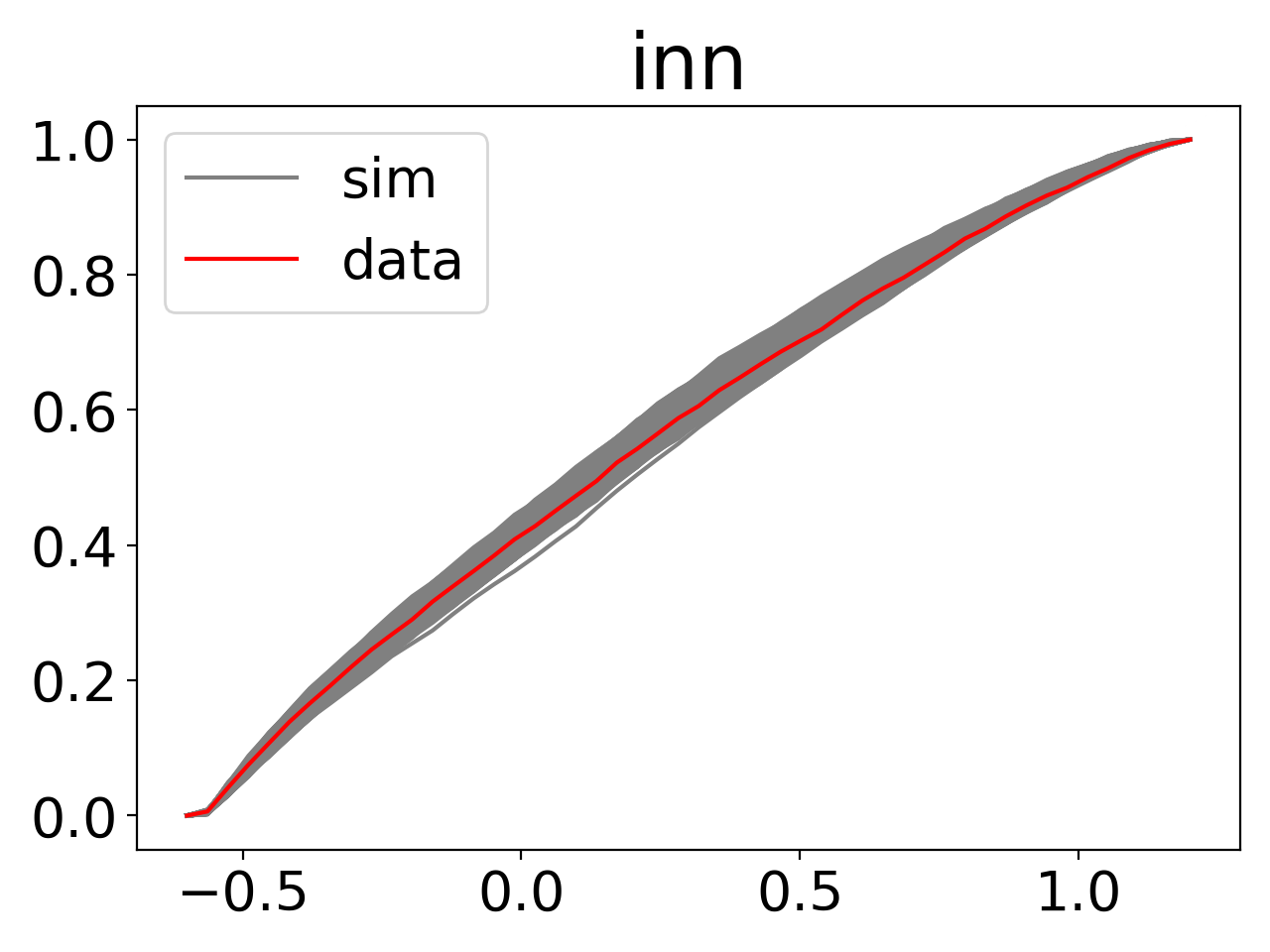}
\end{tabular}
\caption{Empirical distribution functions of various orientation characteristics based on the~data (red curves) and $5\,000$ simulations (grey curves) under each of the~fitted models with densities $f_{\text{noint}}$ (first column),  $f_{\text{int}, w^{(0)}}$ (second column), $f_{\text{int}, w^{(1)}}$ (third column) and  $f_{\text{int}, w^{(2)}}$ (last column).}
\label{fig:DFs_comparison}
\end{figure*}

We now use our methodology for analysing the~NiTi dataset introduced in Section~\ref{sec:data}.
Below we first describe Table~\ref{tab:comp} and Figure~\ref{fig:DFs_comparison} and second comment on our findings.

\begin{table}[tb]
\setlength{\tabcolsep}{8pt} 
\renewcommand{\arraystretch}{1.5} 
\caption{Empirical results for the~NiTi dataset and simulated results under fitted models. First row: The~maximum pseudolikelihood estimate of $\theta$ for the~three pairwise interaction models. Second row: The~corresponding values for the~maximized log-pseudolikelihood function. The~remaining rows: Numerical results for mean values and standard deviations of various orientation characteristics based on Monte Carlo estimates (first to fourth columns) and the~NiTi dataset (last column).}
\label{tab:comp}
\centering
 \begin{tabular}{c|c|ccc|c} 
  & $f_{\text{noint}}$ &  $f_{\text{int}, w^{(0)}}$ &  $f_{\text{int}, w^{(1)}}$ &  $f_{\text{int}, w^{(2)}}$ & data  \\
\hline \hline
 $\hat{\theta}$ & -- & $0.17$ & $0.29$ & $2.82$ & -- \\
 $l(\hat{\theta})$ & -- & $-243.90$ & $-257.06$ & $\mathbf{-240.32}$ & -- \\
\hline \hline
 tilt [mean] & $0.953$ & $\mathbf{0.959}$ & $0.958$ & $\mathbf{0.959}$ & $0.959$ \\
 tilt [sd] & $0.048$ & $\mathbf{0.043}$ & $0.044$ & $\mathbf{0.043}$ & $0.039$ \\
\hline
 $\dis$ [mean] & $24.257$ & $35.213$ & $35.846$ & $\mathbf{35.698}$ & $35.498$ \\
 $\dis$ [sd]  & $14.692$ & $\mathbf{13.660}$ & $13.583$ & $13.608$ & $13.889$ \\
\hline
 $\inn$ [mean]  & $0.565$ & $0.190$ & $0.167$ & $\mathbf{0.173}$ & $0.181$ \\
 $\inn$ [sd] & $0.517$ & $\mathbf{0.489}$ & $0.486$ & $0.487$ & $0.496$\\
\hline
 $\hat{d}$ & $0.634$ & $\mathbf{1.061}$ & $1.071$ & $1.068$ & $1.061$
 \end{tabular}
\end{table}

Table~\ref{tab:comp} summarizes numerical results based on partly pseudolikelihood estimation (first and second rows) and partly selected orientation characteristics (remaining rows) given by the~tilt with choice $v=(1,1,1)^\top$, disorientation angle, inner product and sample dispersion, cf.\ Section~\ref{sec:chars}. The~second row shows the~values of the~maximized log-pseudolikelihood function for the~three pairwise interaction models, where the~largest value is highlighted. 
The last column shows the~results for the~data and the~first four columns concern the~results obtained from $5\,000$ simulations under each of the~four fitted models in Section~\ref{sec:models}. The~closest values to the~last column are highlighted. When simulating from models $f_{\text{int}, w^{(k)}}$, $k=0,1,2$, $1\,000$ sweeps from the~Metropolis-within-Gibbs algorithm were sufficient (here, we omit the~time series plots for various statistics which we considered).

Figure~\ref{fig:DFs_comparison} shows the~empirical distribution functions of various orientation characteristics as calculated from the~data and $5\,000$ simulations under each of the~four fitted models in Section~\ref{sec:models}. The~figure may be divided into the~first four rows, which concern the~samples of orientations, and the~last two rows, which concern the~samples obtained by considering all pairs of neighbouring orientations.

Regarding the~first model with pdf $f_{\text{noint}}$ (see \eqref{eq:p_noint}), considering the~first column in Figure~\ref{fig:DFs_comparison}, we see that the~two first Euler angles are fitted well, but less so for the~third Euler angle and the~tilt. Moreover, the~distribution functions of the~disorientation angle and the~inner product characteristic are far from the~real data.

For the~pairwise interaction model with weight $w^{(0)}$ (and having pdf $f_{\text{int}, w^{(0)}}$, cf.\ \eqref{eq:general}), we compare with the~results above for the~first model as well as the~data.  
Comparing the~two first columns in Figure~\ref{fig:DFs_comparison}, we see not much difference with respect to the~two first Euler angles, but there is a~slight improvement for the~third Euler angle and the~tilt, and as expected a~pronounced improvement for the~disorientation angle and the~inner product characteristic. Moreover, all the~values of the~orientation characteristics in Table~\ref{tab:comp} are now closer to the~corresponding values for the~data. 

For the~pairwise interaction model with weight $w^{(1)}$ (and having pdf $f_{\text{int}, w^{(1)}}$), we see the~following when comparing with the~data and the~two models above.
Based on the~third column in Figure~\ref{fig:DFs_comparison}, in comparison to the~first model (the first column), we again observe a~slight improvement regarding the~Euler angles and the~tilt, as well as an~improvement regarding the~disorientation angle and the~inner product characteristic. On the~other hand, with respect to the~pairwise interaction model with weight $w^{(0)}$ (the second column) we do not observe any change.  
Further, considering next Table~\ref{tab:comp}, all the~values of the~orientation characteristics are closer to the~corresponding values for the~data than for the~first model, but farther away from the~data than for the~pairwise interaction model with weight $w^{(0)}$. In addition, the~value of the~maximized log-pseudolikelihood function is lower. 

Finally, we compare the~pairwise interaction model with weight $w^{(2)}$ (and having pdf $f_{\text{int}, w^{(2)}}$) with the~data and the~three models above.
Considering the~last column in Figure~\ref{fig:DFs_comparison}, in comparison to the~first model we observe again a~slight improvement with respect to the~Euler angles and the~tilt, as well as an~improvement with respect to the~disorientation angle and the~inner product characteristic. However, we do not observe any change with respect to the~pairwise interaction models with weights $w^{(0)}$ and $w^{(1)}$. 
Furthermore, all the~orientation characteristics in Table~\ref{tab:comp} are closer to the~data in comparison to the~first model and to the~pairwise interaction model with weight $w^{(1)}$. On the~other hand, some of the~values are farther away from the~data than for the~pairwise interaction model with weight $w^{(0)}$ and some of the~values are closer to the~data. The~values for the~tilt are the~same.
Moreover, the~value of the~maximized log-pseudolikelihood function is now the~largest. 

In conclusion, we do not obtain a~clear picture from Figure~\ref{fig:DFs_comparison} except that one of the~pairwise interaction models should be preferred, while the~results in Table~\ref{tab:comp} show that one of the~pairwise interaction models with weight $w^{(0)}$ or $w^{(2)}$ should be preferred.

\section*{Acknowledgement} 

We thank the~Czech Science Foundation for their support of the~project no. 22-15763S and the Grant schemes at Charles University, project no. CZ.02.2.69/0.0/0.0/19\_073/0016935.

\appendix

\section{Transversal for a cubic lattice structure}
\label{sec:transversal}

In the case of cubic crystal symmetry, the~subgroup $\mathcal{O}$ has 24 elements and thus each equivalence class of $\mathrm{SO}(3)$ contains 24 symmetrically equivalent orientations. A~transversal is obtained by taking exactly one representative from each equivalence class. When dealing with Euler angles $\left( \varphi_1,\, \phi,\, \varphi_2 \right)$, a~particular partition of $[0,2\pi) \times [0,\pi] \times [0,2\pi)$ into 24 subsets (representing transversals) is described in \cite[Section 2.6.2]{Engler} and \cite[Section 3.4]{Hansen1978}. However, this description lacks the~specification of the~boundaries of the~subsets. The~subset called III in \cite[Section 2.6.2]{Engler} and \cite[Section 3.4]{Hansen1978} corresponds to the~fundamental zone $F$ defined by \eqref{eq:fz} in Section \ref{sec:fundamental}. It is not difficult to see that $F$ is not a~transversal for $\mathcal{O}$ in $\mathrm{SO}(3)$. For example, orientations $G_1$, $G_2$ given by Euler angles $(0,\pi/2,0), (\pi,\pi/2,0) \in F$, respectively, are equivalent because $G_2 = SG_1$ for some $S \in \mathcal{O}$. All symmetries belonging to $\mathcal{O}$ could be described by Euler angles which are multiples of $\pi/2$. In particular, $S$ is represented by $(\pi,\pi,0)$. We use the~notation $(\pi,\pi/2,0) = (\pi,\pi,0) (0,\pi/2,0)$ to rewrite the~relation $G_2 = SG_1$.

\begin{figure}[tb]
\centering
\includegraphics[width=0.5\textwidth]{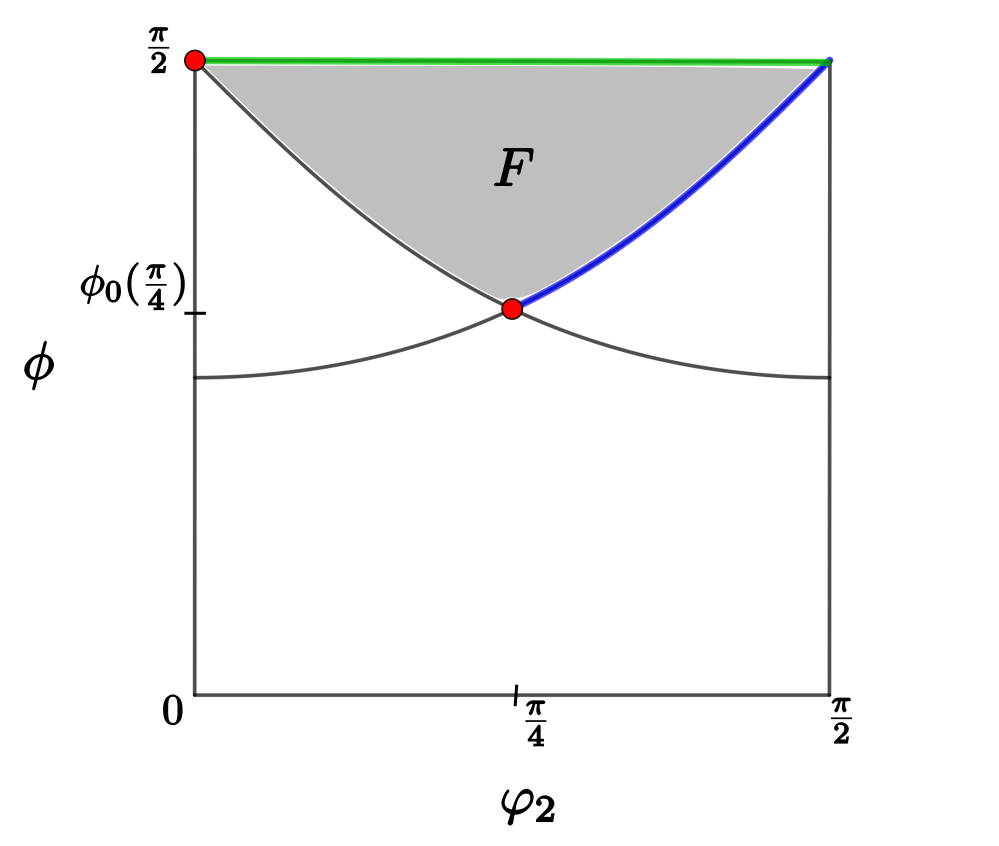}
\caption{A cross-section of the~fundamental zone $F$ given by \eqref{eq:fz}. The~sets of points that need to be considered in order to get the~transversal $F_0$ are highlighted.}
\label{fig:transversal}
\end{figure}

By thoroughly investigating the~boundary of $F$ we find out which orientations to exclude from $F$ in order to obtain a~transversal. In this way, we get that a~transversal $F_0$ contained in $F$ could be taken as
\[
F_0 = F \setminus (F_1 \cup F_2 \cup F_3 \cup F_4),
\]
where
\begin{align*}
F_1 &= \{(\varphi_1,\phi_0(\varphi_2),\varphi_2) \mid \varphi_1 \in [0,2\pi), \varphi_2 \in (\pi/4,\pi/2)\}, \\
F_2 &= \{(\varphi_1,\pi/2,\varphi_2) \mid \varphi_1 \in [\pi,2\pi),
\varphi_2 \in (0,\pi/2)\}, \\
F_3 &= \{(\varphi_1,\phi_0(\pi/4),\pi/4) \mid \varphi_1 \in [2\pi/3,2\pi)\}, \\
F_4 &= \{(\varphi_1,\pi/2,0) \mid \varphi_1 \in [\pi/2,2\pi)\}.
\end{align*}
Note that $F$ and $F_0$ have the~same interiors and $\mu(F \setminus F_0) = 0$.
The explanation of the~form of individual sets that are excluded from $F$ follows below. For this purpose it is useful to observe the~contour line of $F$ with given $\varphi_1$ shown in Figure \ref{fig:transversal}.

First consider $(\varphi_1,\phi_0(\varphi_2),\varphi_2)$ with $\varphi_2 \in (0,\pi/4)$. This orientation is equivalent to\linebreak
$(\tilde{\varphi_1},\phi_0(\varphi_2),\pi/2-\varphi_2)$ with $\tilde{\varphi_1}$ possibly different from $\varphi_1$. 
Therefore, we restrict only to $\varphi_2 \in (0,\pi/4)$ in $F_0$
and remove the~set $F_1$, where $\varphi_2 \in (\pi/4,\pi/2)$, 
depicted by a~blue colour in Figure \ref{fig:transversal}.
The symmetry that transforms $(\varphi_1,\phi_0(\varphi_2),\varphi_2)$ to $(\tilde{\varphi_1},\phi_0(\varphi_2),\pi/2-\varphi_2) \in F_1$ is given by Euler angles $(\pi/2,\pi/2,0)$. It means that 
\[
\mbox{$(\tilde{\varphi_1},\phi_0(\varphi_2),\pi/2-\varphi_2) =
(\pi/2,\pi/2,0) (\varphi_1,\phi_0(\varphi_2),\varphi_2)$ 
for $\varphi_2 \in (0,\pi/4)$.}
\]

The form of $F_2$ (green segment in Figure \ref{fig:transversal}) follows from the~relation 
\[
\mbox{$(\varphi_1+\pi,\pi/2,\pi/2-\varphi_2) = (\pi/2,\pi,0) (\varphi_1,\pi/2,\varphi_2)$
for $\varphi_1 \in [0,\pi)$ and $\varphi_2 \in (0,\pi/2)$.
}
\] 
It means that for $\phi=\pi/2$ we exclude one half of the~$\varphi_1$ values from $F$.

If $\phi=\phi_0(\pi/4)=\arccos (1/\sqrt{3})$ and $\varphi_2 = \pi/4$
(right bottom red dot in Figure \ref{fig:transversal})
we exclude two thirds of the~$\varphi_1$-range (set $F_3$) because
\[
(\varphi_1+2\pi/3,\phi_0(\pi/4),\pi/4) = (\pi/2,\pi/2,0) (\varphi_1,\phi_0(\pi/4),\pi/4)\]
and
\[
(\varphi_1+4\pi/3,\phi_0(\pi/4),\pi/4) = (\pi,\pi/2,\pi/2) (\varphi_1,\phi_0(\pi/4),\pi/4)
\]
for $\varphi_1 \in [0,2\pi/3)$. 

Finally, similar reasoning applies to the~case $\phi=\pi/2$ and $\varphi_2=0$ (left upper red dot in Figure \ref{fig:transversal}). Then $(\varphi_1,\pi/2,0)$ with $\varphi_1 \in [0,\pi/2)$ is equivalent to $(\varphi_1+k\pi/2,\pi/2,0)$, $k=1,2,3$. More specifically, we have
\begin{align*}
(\varphi_1+\pi/2,\pi/2,0) &= (\pi/2,\pi/2,3\pi/2) (\varphi_1,\pi/2,0), \\
(\varphi_1+\pi,\pi/2,0) &= (\pi,\pi,0) (\varphi_1,\pi/2,0), \\
(\varphi_1+3\pi/2,\pi/2,0) &= (3\pi/2,\pi/2,\pi/2) (\varphi_1,\pi/2,0)
\end{align*}
for $\varphi_1 \in [0,\pi/2)$. Any transversal must contain only one of four equivalent orientations $(\varphi_1+k\pi/2,\pi/2,0)$, $k=0,1,2,3$. We choose $\varphi_1 \in [0,\pi/2)$ in $F_0$.

\section{Derivatives of the~log-pseudolikelihood function}
\label{sec:MPL}

The log-pseudolikelihood function in Section~\ref{sec:methods} is given as
\begin{align*}
l (\theta) & = \sum_{i=1}^n\ln f_{i\theta}(g_i\mid (\g_n)_{-i},\C_n) \\
& = \sum_{i=1}^n \ln \left(\frac{1}{c_{i}(\theta, \C_n)} f_{\rm s}(g_i) \exp \left\{ \theta \sum_{j:\,i \sim j} w_{ij} \inn(g_i, g_j) \right\}\right) \\
& =  \sum_{i=1}^n \left( - \ln c_{i}(\theta, \C_n) + \ln f_{\rm s}(g_i) + \theta \sum_{j:\,i \sim j} w_{ij} \inn(g_i, g_j) \right).
\end{align*}
The first and second derivatives of $l(\theta)$ with respect to the~parameter $\theta$ are
\[
l'(\theta) = \sum_{i=1}^n \left( - \frac{c_{i}'(\theta, \C_n)}{c_{i}(\theta, \C_n)}+ \sum_{j:\,i \sim j} w_{ij} \inn(g_i, g_j) \right)
\]
and 
\[
l'' (\theta) = \sum_{i=1}^n \left( \frac{c_{i}'(\theta, \C_n)}{c_{i}(\theta, \C_n)} \right)^2 - \sum_{i=1}^n \frac{ c_{i}''(\theta, \C_n) }{ c_{i}(\theta, \C_n)} .
\]

Consider $t(F)$ (see Section \ref{sec:quotient}) equipped with the~measure \eqref{eq:Leb}. It is easily seen that $t(F)$ has volume $\pi^2/3$. Divide $t(F)$ into $M$ equally large cells with midpoints $\left\{ u_m, \, m = 1,\dots,M \right\}$. Denoting $t^{-1}: (\varphi_1, \eta, \varphi_2) \mapsto (\varphi_1, \phi, \varphi_2)$, we approximate the~normalizing constant
\begin{align*}
c_{i}(\theta, \C_n)  & = \int f_{\mathrm s}(g) \exp \left\{ \theta \sum_{j:\,i\sim j}w_{ij} \inn(g,g_j) \right\} \, \mathrm d \mu_F(g) \\
& = \int_{t(F)} f_{\mathrm s}(t^{-1}(\varphi_1, \eta, \varphi_2)) \ \times \\
& \quad \quad \times \exp \left\{ \theta \sum_{j:\,i\sim j}w_{ij} \inn(t^{-1}(\varphi_1, \eta, \varphi_2), (\varphi_{1j}, \phi_j, \varphi_{2j})) \right\} \, \mathrm d(\varphi_1, \eta, \varphi_2) \\
& \approx \frac{\pi^2}{3M} \sum_{m=1}^M f_{\mathrm s}(t^{-1}(u_m)) \exp \left\{\theta\sum_{j:\,i\sim j} w_{ij} \inn(t^{-1}(u_m),g_j)\right\}.
\end{align*}
The approximation of the~first and second derivatives of the~log-pseudolikelihood function with respect to the~parameter $\theta$ are then
\begin{align*}
l'(\theta) \approx & \sum_{i=1}^n \sum_{j: \,i \sim j} w_{ij} \inn(g_i, g_j)  \\
& - \sum_{i=1}^n \dfrac{\sum\limits_{m=1}^M f_{\mathrm s}(t^{-1}(u_m)) \left( \sum\limits_{j:\,i\sim j} w_{ij} \inn(t^{-1}(u_m),g_j) \right) \exp \left\{\theta\sum\limits_{j:\,i\sim j} w_{ij} \inn(t^{-1}(u_m),g_j)\right\}  }{\sum\limits_{m=1}^M f_{\mathrm s}(t^{-1}(u_m)) \exp \left\{ \theta\sum\limits_{j:\,i\sim j} w_{ij} \inn(t^{-1}(u_m),g_j) \right\}} 
\end{align*}
and
\begin{align*}
l'' (\theta) \approx & \sum\limits_{i=1}^n \left( \dfrac{ \sum\limits_{m=1}^M f_{\mathrm s}(t^{-1}(u_m)) \left(\sum\limits_{j:\,i\sim j} w_{ij} \inn(t^{-1}(u_m),g_j) \right) \exp \left\{\theta\sum\limits_{j:\,i\sim j} w_{ij} \inn(t^{-1}(u_m),g_j)\right\} }{\sum\limits_{m=1}^M f_{\mathrm s}(t^{-1}(u_m)) \exp \left\{\theta\sum\limits_{j:\,i\sim j} w_{ij} \inn(t^{-1}(u_m),g_j)\right\} } \right)^2\\
& - \sum_{i=1}^n \dfrac{ \sum\limits_{m=1}^M f_{\mathrm s}(t^{-1}(u_m)) \left( \sum\limits_{j:\,i\sim j} w_{ij} \inn(t^{-1}(u_m),g_j)  \right)^2  \exp \left\{\theta \sum\limits_{j:\,i\sim j} w_{ij} \inn(t^{-1}(u_m),g_j)\right\} }{ \sum\limits_{m=1}^M f_{\mathrm s}(t^{-1}(u_m)) \exp \left\{\theta \sum\limits_{j:\,i\sim j} w_{ij} \inn(t^{-1}(u_m),g_j)\right\} }.
\end{align*}

\bibliographystyle{spmpsci} 
\bibliography{lit_lattice}

\end{document}